\documentclass[fleqn,usenatbib]{mnras}
\usepackage{newtxtext,newtxmath}

\usepackage{newtxtext,newtxmath}

\usepackage[T1]{fontenc}

\DeclareRobustCommand{\VAN}[3]{#2}
\let\VANthebibliography\thebibliography
\def\thebibliography{\DeclareRobustCommand{\VAN}[3]{##3}\VANthebibliography}

\usepackage{graphicx}	
\usepackage{amsmath}	

\usepackage{hyperref}
\usepackage{booktabs}
\usepackage{amsmath}
\usepackage{ulem}
\usepackage{footmisc} 
\usepackage{color}

\title[A novel approach to determining the age of stellar clusters]{A novel approach to determining the age of stellar clusters: application to NGC 188.}

\author[K. Yakut, B. Kalomeni and S. Rappaport]{
Kadri Yakut,$^{1,2}$\thanks{E-mail: kadri.yakut@ege.edu.tr (KY)}
Belinda Kalomeni$^{1}$
and Saul Rappaport$^{3}$
\\
$^{1}$Department of Astronomy and Space Sciences, Faculty of Science, Ege University, 35100, {\.I}zmir, T\"urkiye\\
$^{2}$Institute of Astronomy, The Observatories, Madingley Road, Cambridge CB3 OHA, UK\\
$^{3}$Department of Physics, Kavli Institute for Astrophysics and Space Research, M.I.T., Cambridge, USA}

\date{Accepted XXX. Received YYY; in original form ZZZ}

\pubyear{\the\year{}}

\begin{document}
\label{firstpage}
\pagerange{\pageref{firstpage}--\pageref{lastpage}}
\maketitle

\begin{abstract}
We present a new independent determination for the age of one of the oldest open clusters, NGC 188: $6.41 \pm 0.33$ Gyr. We utilize a novel approach where we jointly fit the radial velocities (RVs) and spectral energy distributions (SED) of six binary star systems in the cluster. The joint fit has 21 free parameters: 12 stellar masses, 6 orbital inclination angles, as well as a common cluster age, distance, and extinction, $A_V$. The fit to the SEDs utilizes MIST stellar isochrones, and the fitted stellar parameters are presented in HR diagrams (R -- $T_{\rm eff}$, R -- M, and M -- $T_{\rm eff}$) showing the stars’ different states of evolution. These isochrones are compared with PARSEC and Y2 isochrones to obtain an estimate of the uncertainties introduced by different stellar models. Of the 3017 stars taken from the NGC 188 field, 333 possible member stars were selected using Gaia data and analyzed separately for their mean distances and proper motions. We find a distance to NGC 188 of $1850 \pm 12$ pc compared with the independent photometric distance found from the SED fitting of $1897 \pm 58$ pc.  
\end{abstract}

\begin{keywords}
Binary stars --- Fundamental parameters of stars ---  (Galaxy:) open clusters and associations: individual (NGC 188) --- (Galaxy:) open clusters and associations: general ---  proper motions
\end{keywords}



\section{Introduction} 
\label{sec:intro}
Open clusters, containing stars of nearly identical ages and chemical compositions, serve as essential laboratories for testing  stellar evolution and dynamical models. Near-main-sequence binaries, particularly eclipsing and double-lined spectroscopic systems (EBs and SB2s), are valuable tools in the ongoing effort to constrain cluster ages and stellar parameters via joint light curve and radial velocity analyses \citep{Meynet1993,Southworth2004,Geller2015,Kroupa2001,Yakut2003}. These binaries also provide insights into mass transfer histories and angular momentum evolution \citep{Rappaport1983,Pols1998,Hurley2004,Yakut2005,Eggleton2017}, making them valuable for probing stellar physics and the internal dynamics of open clusters \citep{Mardling2001,Yakut2015}. Recent advances from the Gaia mission \citep{GaiaCollaboration2018,GaiaCollaboration2021} have revolutionized cluster studies by providing microarcsecond precision in astrometric parameters, enabling robust membership determinations, and reducing field star contamination. The integration of Gaia astrometry with binary analyses can improve the accuracy of cluster properties and provide rigorous tests of stellar evolution and dynamical models.

The open cluster NGC 188 is one of the oldest and most studied open clusters in the Milky Way. Its proximity, abundance, and large Galactic latitude make it an ideal target for studying long-term stellar and dynamical evolution. In addition to NGC 188, other old open clusters in the Galaxy include Berkeley 17 \citep{Phelps1994}, NGC 6791 \citep{King2005,Yakut2015}, Berkeley 39 \citep{Bragaglia2012}, and M67 \citep{Sarajedini2009,Yakut2009}. Various methods have been used to determine the age of NGC 188, including isochrone fitting to color-magnitude diagrams, analysis of eclipsing binaries, white dwarf cooling sequences, and stellar population synthesis \citep{Cohen2020,Bossini2019}. Recent studies using Gaia DR3 data have further refined these estimates by allowing more accurate cluster selection and astrometric filtering \citep{Randich2018,CantatGaudin2020,Dursun2024}. The derived ages for NGC~188 are typically in the range of 6 to 8 Gyr, although some studies suggest slightly older values depending on model assumptions and metallicity calibrations.

Studies of NGC 188 and similarly aged clusters are essential for understanding the mechanisms that govern their long-term survival in the Galactic disk. The dynamical evolution of open clusters is driven by external tidal forces as well as by internal processes such as mass segregation and the energy input from hard binaries. Detailed spectroscopic surveys have shown that hard binaries act as significant energy sources, maintaining the dynamical equilibrium of clusters on Gyr timescales \citep{Geller2009}. Analyses of binary populations further reveal their critical role in shaping the global structure and survivability of the cluster \citep{Meibom2009,Geller2015}. The combination of deep photometric surveys, high-resolution spectroscopy, and Gaia astrometry has established NGC 188 as an indispensable benchmark for probing stellar evolution and cluster dynamics at late ages.

NGC 188 is one of the best chemically characterized old open clusters, providing valuable insights into stellar and Galactic chemical evolution. High-resolution spectroscopic studies consistently report iron abundances close to the Sun, with [Fe/H] values between +0.00 and +0.12 dex \citep{Friel2010,Sun2022}. Elemental abundance ratios, including $\alpha$-elements and iron-peak elements, show homogeneity consistent with thin disk chemical patterns \citep{Friel2010}. Compared to M67 - a solar metallicity cluster - NGC 188 has a slightly super-solar iron content, reinforcing its role as a benchmark for tracing chemical evolution beyond the solar age. Recent age and metallicity determinations for NGC 188 are summarized in Table~\ref{tab:ngc188_metallicity}.

In this work, we present a novel approach to determining the age of open clusters, and focus on the very old cluster NGC 188.  In particular we carry out a joint, i.e., simultaneous, fit to the radial velocities and spectral energy distributions of six double-lined spectroscopic binary systems, utilizing stellar evolution models.  As we demonstrate, this type of analysis does not require eclipsing binaries.  One of the free parameters is the age of the system, which emerges with both a statistical uncertainty as well as one for the uncertainty in the cluster metallicity. In Sec~\ref{sec:obs}, we describe the selection criteria for the target systems, highlighting their suitability based on radial velocity signatures, absence of mass transfer, and brightness thresholds. In Sec~\ref{sec:lcrvmodel}, we describe the combined light curve and radial velocity modeling for the two eclipsing binaries PKM~4705 and PKM~5762, using TESS photometric and spectroscopic data to derive precise orbital and stellar parameters. In Sec~\ref{sec:SEDs}, we perform a joint spectral energy distribution (SED) analysis for all six binaries, incorporating Gaia astrometry and using MIST stellar evolution models to assess the evolutionary states of the component stars and derive the cluster's age and distance. Sec~\ref{sec:astrometric} presents the astrometric membership analysis using Gaia DR3 proper motions and parallaxes, which refines the cluster membership and provides additional constraints on the global cluster parameters. Finally, in Sec~\ref{sec:results} we summarize the results of the photometric, spectroscopic, and astrometric analyses to provide an updated and precise characterization of the fundamental properties of NGC 188.

\begin{table}
\centering
\caption{[Fe/H] and age estimates for the NGC 188 cluster from some studies during the last quarter century.}
\label{tab:ngc188_metallicity}
\begin{tabular}{rrl}
\hline
[Fe/H] & Age (Gyr) & Reference\\
\hline
$-0.12 \pm 0.16$   & 7.7   & \cite{Hobbs_Thorburn_1990}\\
$+0.075 \pm 0.045$ & –     & \cite{Worthey_Jowett_2003}\\
$+0.01 \pm 0.08$   & –     & \cite{Randich_2003} \\
$+0.12 \pm 0.02$   & 7.2   & \cite{Friel_2010} \\
$+0.125 \pm 0.003$ & 5.78  & \cite{Hills_2015} \\
$-0.077 \pm 0.003$ & 6.45  &  \\
$+0.064 \pm 0.018$ & 6.3   & \cite{Sun_2022} \\
$-0.03 \pm 0.07$   & 7.1   & \cite{Carbajo-Hijarrubia2024}\\
$+0.018 \pm 0.005$ & 6.17  & \cite{Childs2024}\\
\hline
Formal average  & & \\
$+0.025 \pm 0.028$ & {$6.41 \pm 0.33$} &  This work\\
\hline
\end{tabular}
\end{table}

\section{Observations}
\label{sec:obs}

\begin{table*}
\scriptsize
\caption{Basic parameters for six binary systems.}
\label{tab:basic_parameters_all}
\begin{tabular}{l r r r r r r}
\hline
                                & PKM~5762  &  PKM~4705         & PKM~4986          &  PKM~4506         & PKM~5147        & PKM~4411  \\
\hline
TIC ID                          & 461618602         & 461601177         & 461601327         & 461601456          & 461601112           & 461601084  \\
Gaia DR3 ID                     & 573937274335905152& 573941053907094144& 573941672382352256& 573944215002950144 & 573940675950545920  & 573935350190553216 \\
2MASS ID                        & J00523772+8510346 & J00452260+8512381 & J00464319+8516056 & J00443634+8521098  & J00465521+8510149   & J00440170+8509063  \\
Alias                           & V12, SMV 10382    & NGC~188~1116, V11 & NGC 188 1040      & NGC 188 3129       & NGC 188 2188        & NGC 188 3019   \\
                                & DGV 407, V785 Cep & SMV 6943, DGV 643 & DGV 738           & DGV 1432           & DGV 563             & DGV 551 \\
$\alpha$ ($^{\rm o}$)           & 13.1569           & 11.3440           & 11.6799           & 11.1515            & 11.7299             & 11.0069 \\
$\delta$ ($^{\circ}$)           & 85.1763           & 85.2106           & 85.2682           & 85.3528            & 85.1708             & 85.1518 \\
$\mu_{\alpha}$ (mas~yr$^{-1}$)  & -2.2675           & -2.3001           & -2.4313           & -2.2675            & -2.2301             & -2.4051 \\
$\mu_{\delta}$ (mas~yr$^{-1}$)  & -0.8240           & -0.9817           & -1.0482           & -0.8240            & -0.9595             & -1.0860 \\
$\varpi$ (mas)                  & 0.5076            & 0.5009            & 0.5083            & 0.5076             & 0.4755              & 0.5036 \\
$G$ (m)                       & 14.599            & 13.770            & 15.124            & 14.599             & 15.188              & 14.510 \\
$G_{\rm{BP}}-G_{\rm{RP}}$ (mag) & 0.888             & 1.172             & 0.874             & 0.888              & 0.907               & 0.956 \\
$V$ (m)                       & 14.76             & 13.84             & 14.76             & 15.27              & 15.34               & 15.69 \\
$B-V$ (m)                     & 0.67              & 0.95              & 0.67              & 0.67               & 0.70                & 0.73 \\
$U-B$ (m)                     & 0.12              & 0.56              & 0.12              & 0.23               & 0.19                & 0.22 \\
\hline
\end{tabular}
\end{table*}

As one of the oldest known open clusters in the Milky Way, NGC 188 has been extensively observed across a wide range of wavelengths under favorable observing conditions. In this study, we selected member stars of NGC 188 that (i) exhibit double-lined radial velocity signatures, (ii) show no evidence of past mass transfer, (iii) are not classified as blue stragglers, and (iv) satisfy a certain brightness threshold. Based on these criteria, six binary systems—PKM 4705, 5762, 4986, 4506, 5147, and 4411—with orbital periods ranging from 6.5 to 48.9 days were identified and simultaneously analyzed to determine both their fundamental stellar parameters and some of the global properties of the cluster.

\begin{figure*}
\centering
\includegraphics[width=0.31\linewidth]{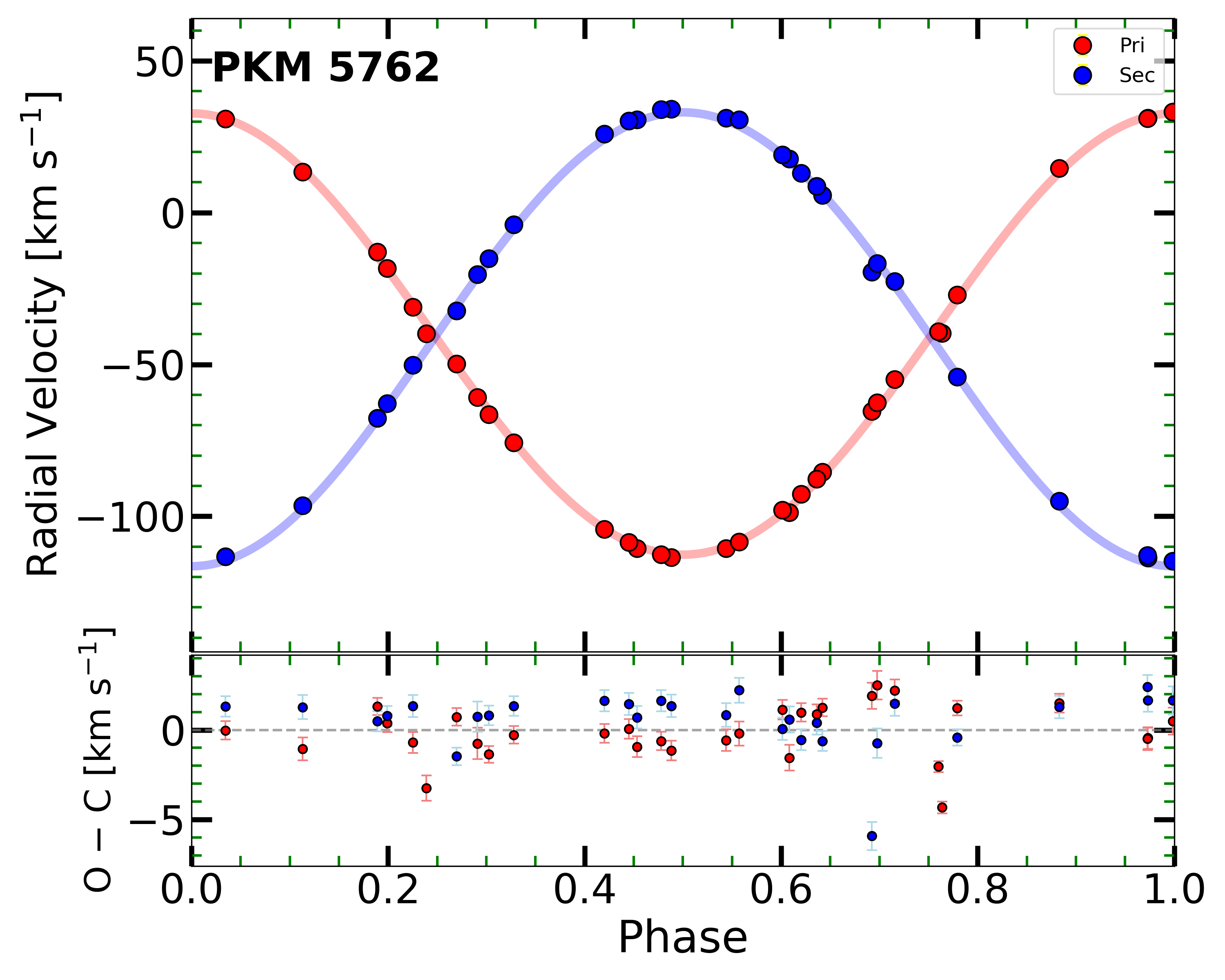}
\includegraphics[width=0.31\linewidth]{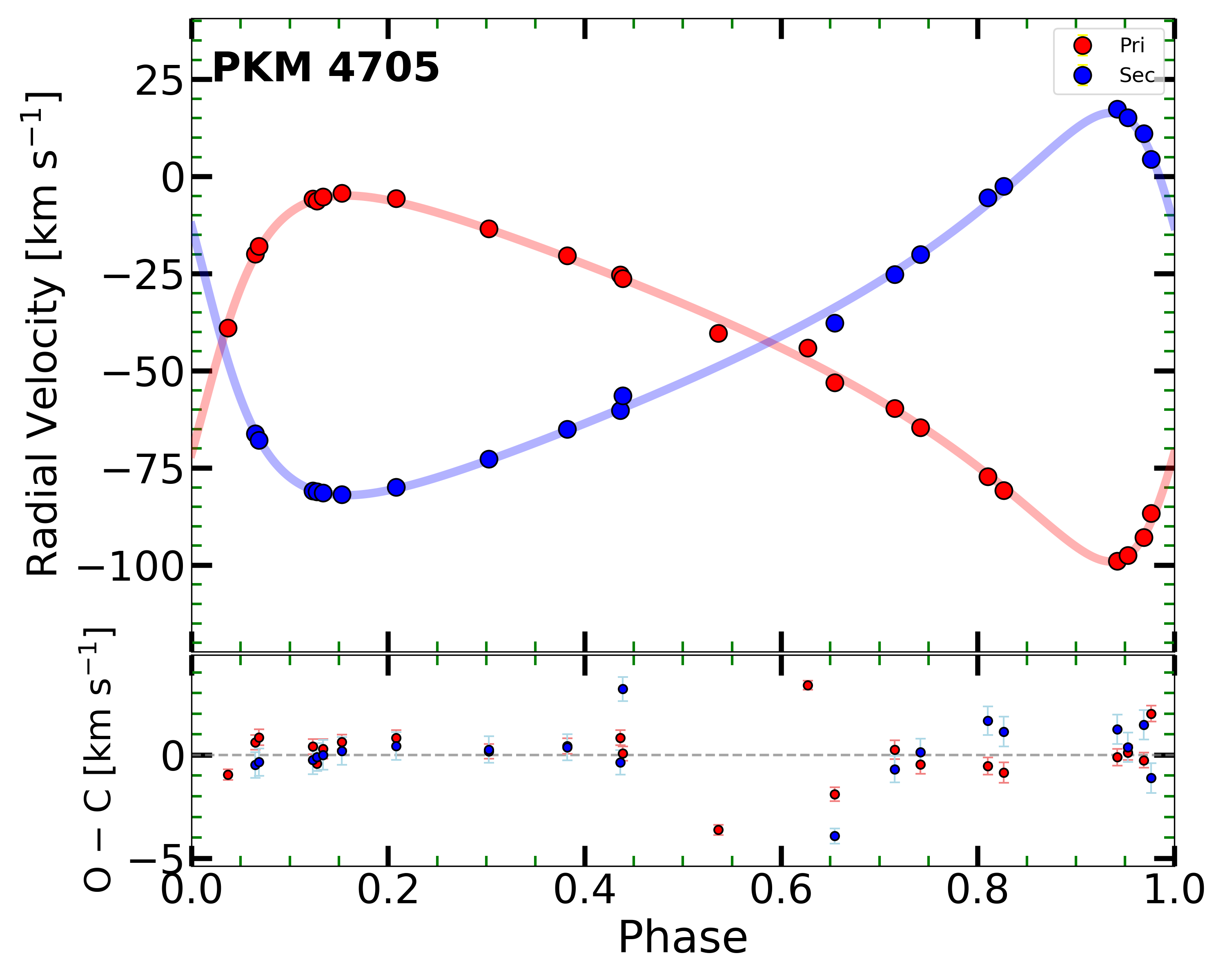}
\includegraphics[width=0.31\linewidth]{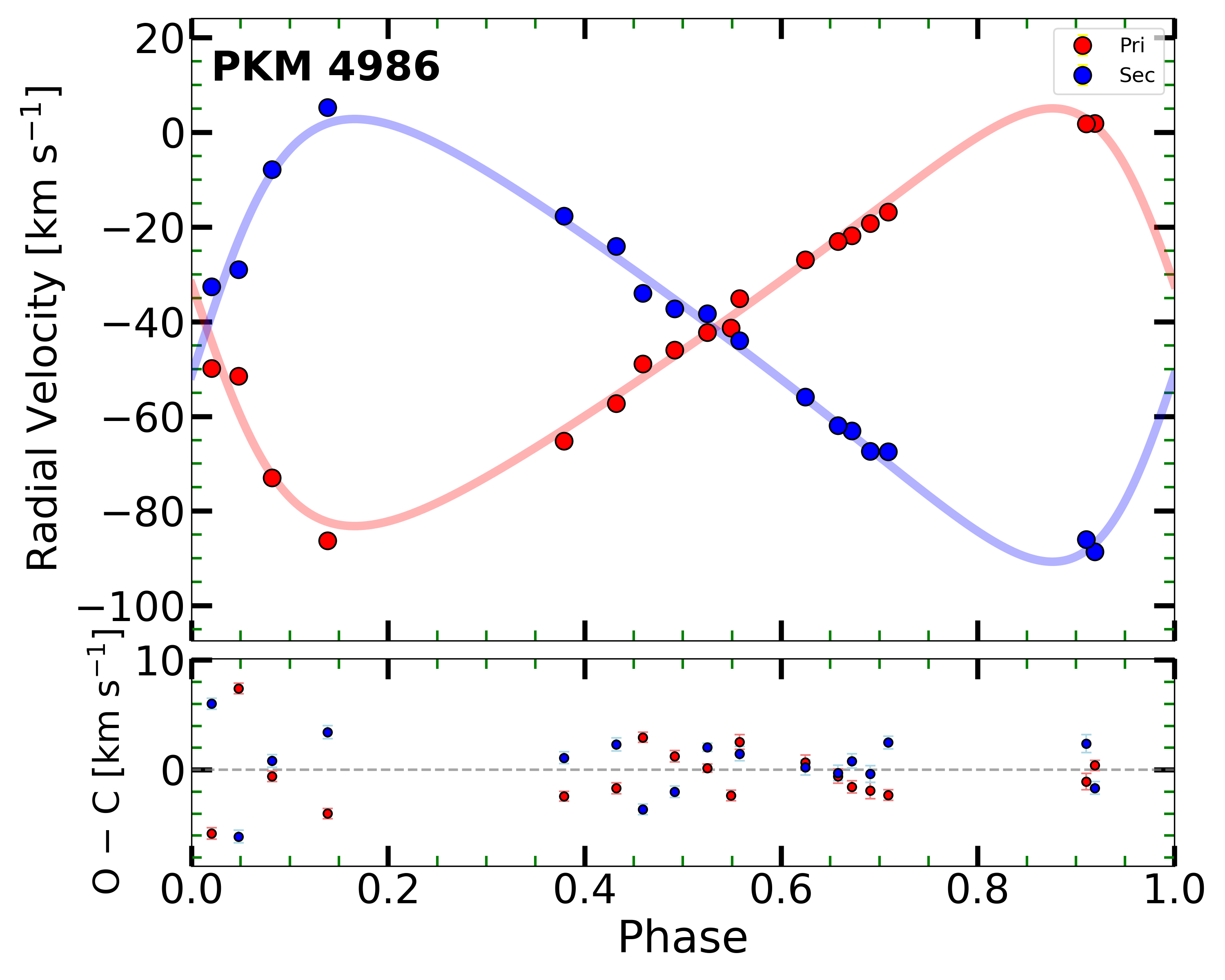}\\
\includegraphics[width=0.31\linewidth]{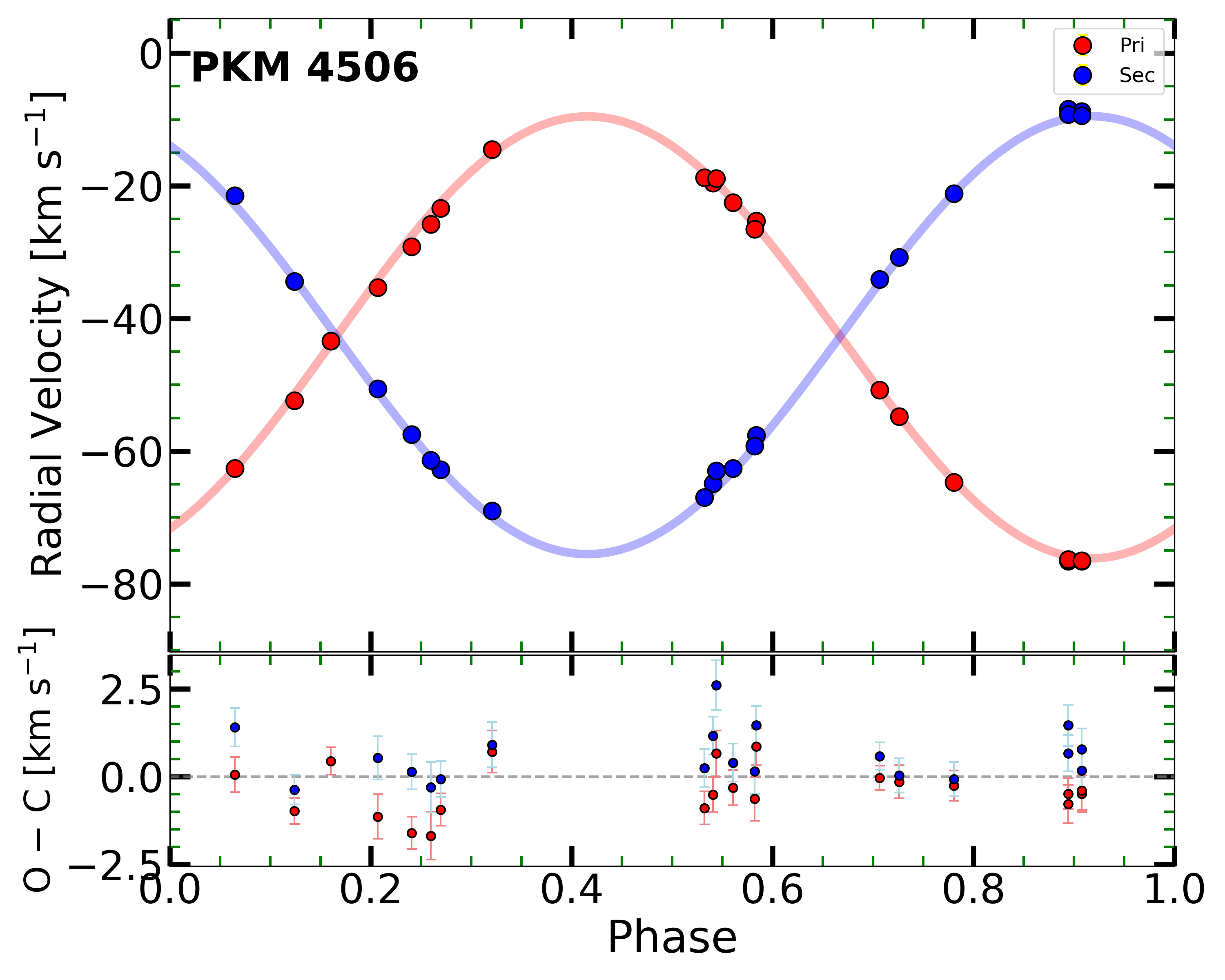}
\includegraphics[width=0.31\linewidth]{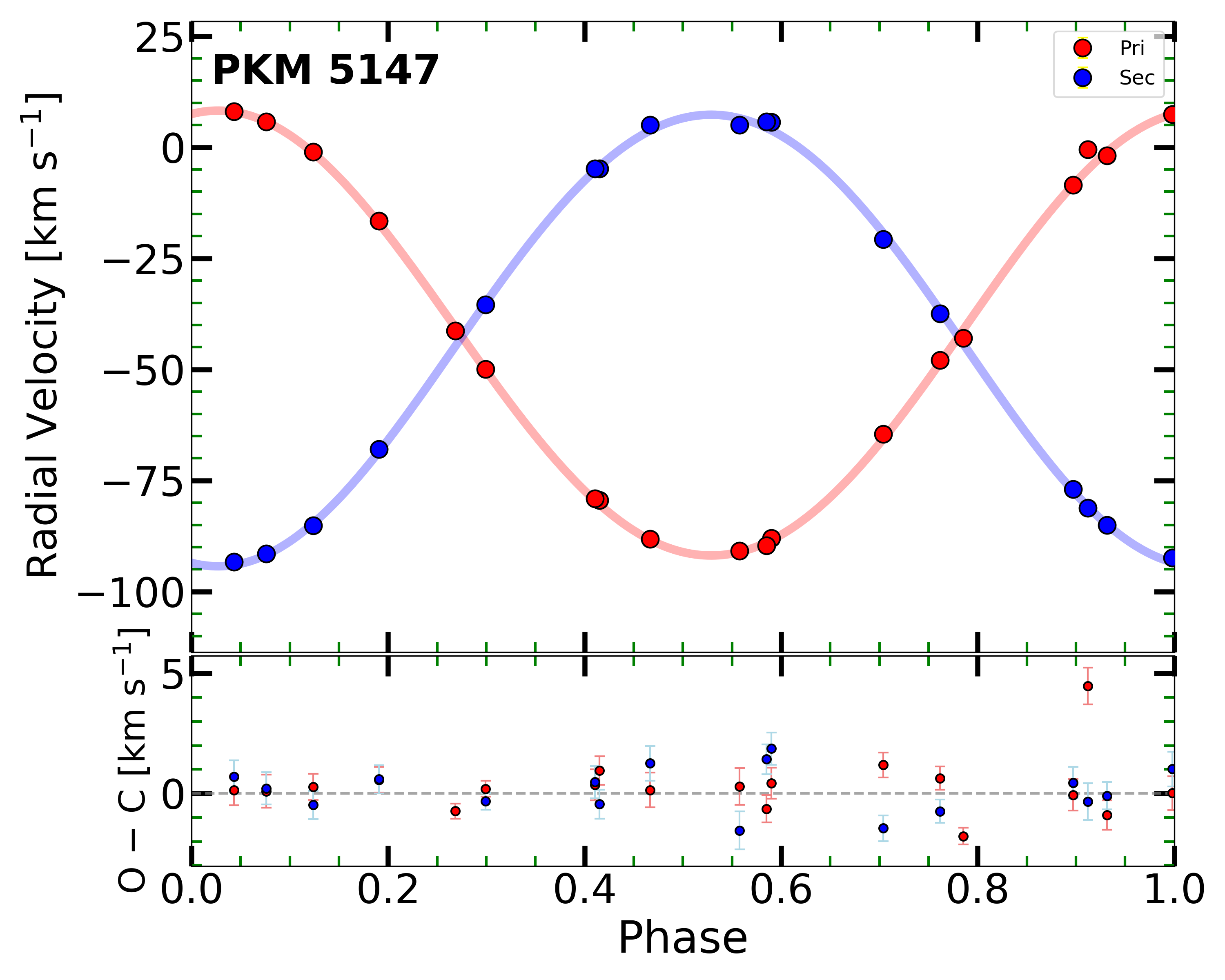}
\includegraphics[width=0.31\linewidth]{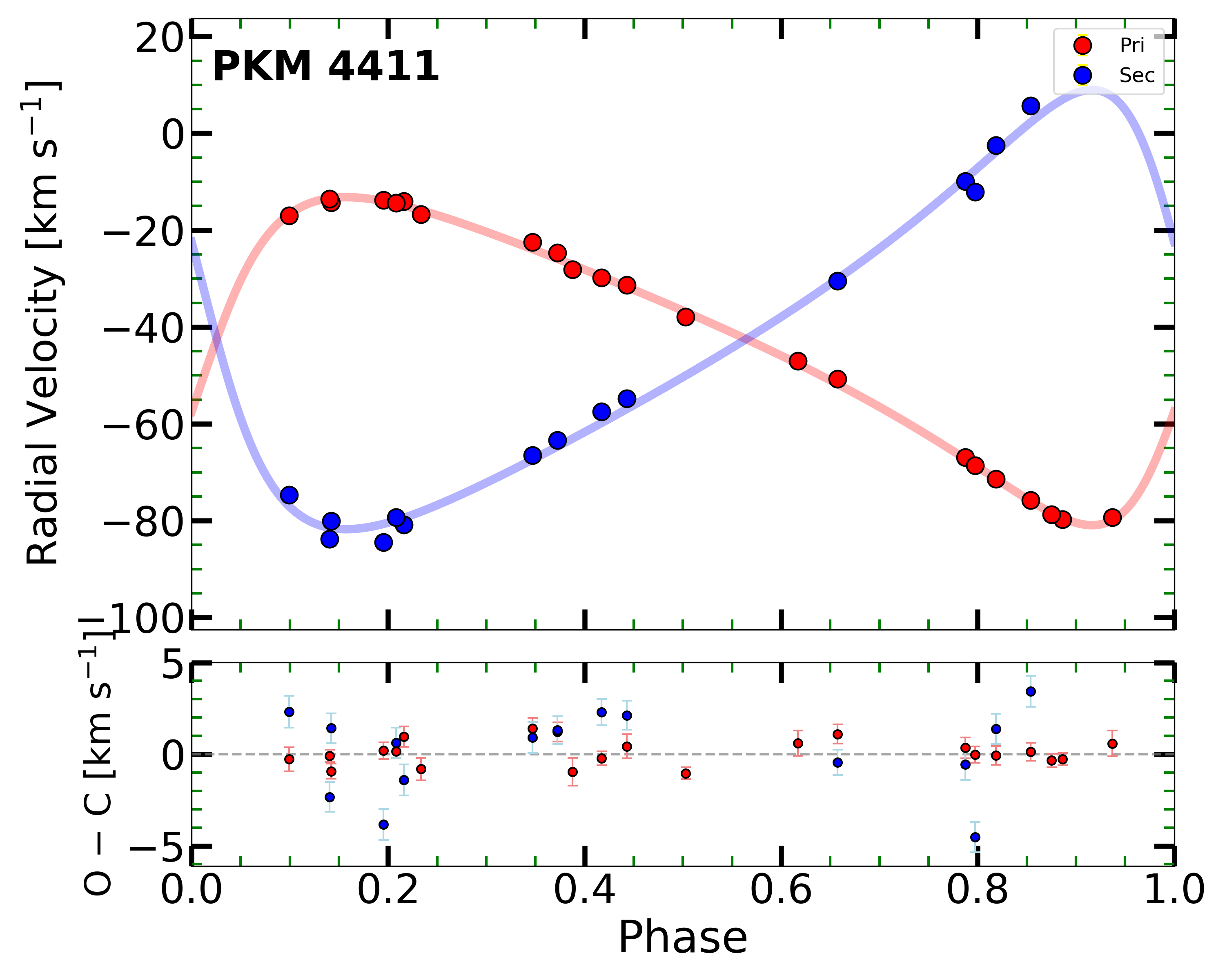}
\caption{Observed and modelled radial velocity variations for all binary systems analyzed in this study, along with the residuals between the observed and fitted velocities. Each panel corresponds to one of the six selected systems in NGC 188. Red symbols represent the primary components, and blue symbols represent the secondary components of the systems. \textit{See text for further details.}}
\label{fig:RVs}
\end{figure*}

As part of the WIYN Open Cluster Radial Velocity Survey, \cite{Geller2009} obtained radial velocity measurements for a large number of binary systems in the open cluster NGC 188. In their study they observed 98 systems and identified 15 of them as double-lined spectroscopic binaries (SB2s). The authors published the full set of radial velocity data they collected, and these measurements serve as the primary source of radial velocity observations for the binary systems analysed in the present study. Figure~\ref{fig:RVs} displays the radial velocity curves for all primary (red symbols) and secondary (blue symbols) components of the binary systems analyzed in this study. In addition to these radial velocity observations, the cluster and many of its member stars have been targeted by numerous large-scale sky surveys, including TESS, Gaia, 2MASS, SDSS, Pan-STARRS, WISE, GALEX, and Swift's UVOT (the references for these data are provided in Section 4). These survey data, supplemented with ground-based observations, were collectively analyzed to fulfill the objectives of this study. The key observational characteristics of the selected binary systems are presented in Table~\ref{tab:basic_parameters_all}.

The Transiting Exoplanet Survey Satellite (TESS) observations were individually examined for each system. PKM 5762 (V785 Cep) was confirmed to be an eclipsing binary, displaying both primary and secondary minima, in agreement with previous studies (e.g., \citealt{Meibom2009}). Conversely, PKM 4705 was identified for the first time as a system exhibiting a primary eclipse, although no secondary minimum was detected. The remaining systems exhibited modulations near maxima, likely attributable to stellar activity, but showed no evidence of eclipses. Consequently, only PKM 5762 and PKM 4705 were subjected to a joint light curve and radial velocity analysis. However, for all six systems, SED analyses were conducted in combination with the radial velocity results. The results of these complementary analyses are presented in the following sections.

\section{Light and radial velocity curves modeling }
\label{sec:lcrvmodel}

For the determination of both orbital and stellar parameters, the precision of the observational data sets is very important. The radial velocities of the PKM 5762 and PKM 4705 binary systems have been successfully obtained by \citep{Geller2008}. In addition to these radial velocities, very accurate photometric datasets obtained by the TESS satellite were used simultaneously in the synthetic model analyses. The double-lined and eclipsing nature of both systems enabled precise determinations of fundamental parameters.
We analyzed the light and radial velocity curves of the eclipsing binary systems PKM 5762 (V12) and PKM 4705 (V11) using the {\sc Phoebe} program \citep{2005ApJ...628..426P}. {\sc Phoebe} is based on the Wilson-Devinney (W-D) methodology \citep{wilson1971ApJ...166..605W,wilson1979ApJ...234.1054W} and implements a differential corrections (DC) algorithm incorporating the Levenberg-Marquardt (L-M) minimization scheme to determine system parameters from observational data.

\begin{table*}
\centering
\caption{Orbital parameters of the six double-lined spectroscopic binary systems analyzed in this study. For each system, the following parameters are listed: $K_1$ and $K_2$ are the radial velocity semi-amplitudes of the primary and secondary components, $P$ is the orbital period, $e$ is the orbital eccentricity, $\omega$ is the argument of periastron, $T_0$ is the time of periastron passage, and $\gamma$ is the systemic velocity}
\label{tab:binaries}
\begin{tabular}{lccccccc}
\hline
ID      & $T_0$           & $P_{\text{orb}}$ & $e$ & $\omega$ & $V_{\gamma}$ & $K_1$        & $K_2$ \\
 PKM    & (HJD - 2400000) & (days)           &     & (deg)    & (km s$^{-1}$)& (km s$^{-1}$)& (km s$^{-1}$) \\
\hline
5762 & $51060.497 \pm 0.001$ & $6.50425  \pm 0.00001$ & $0.012 \pm 0.001$ & $0  $ & $-40.82 \pm 0.08$ & $72.68 \pm 0.15$ & $74.75 \pm 0.17$ \\
4705 & $50752.131 \pm 0.022$ & $35.1836  \pm 0.0016$ & $0.486 \pm 0.002$ & $246 $ & $-42.62 \pm 0.07$ & $47.08 \pm 0.12$ & $49.18 \pm 0.21$ \\
4986 & $53568.893 \pm 0.041$ & $30.08636 \pm 0.00062$ & $0.340 \pm 0.003$ & $81  $ & $-41.44 \pm 0.09$ & $44.13 \pm 0.24$ & $46.78 \pm 0.29$ \\
4506 & $50943.686 \pm 0.005$ & $8.85370  \pm 0.00018$ & $0.006 \pm 0.003$ & $210 $ & $-42.69 \pm 0.08$ & $33.32 \pm 0.17$ & $33.01 \pm 0.19$ \\
5147 & $54356.297 \pm 0.003$ & $6.74296  \pm 0.00015$ & $0.017 \pm 0.003$ & $350 $ & $-42.64 \pm 0.10$ & $50.06 \pm 0.20$ & $50.81 \pm 0.20$ \\
4411 & $50888.797 \pm 0.083$ & $48.8870  \pm 0.0153$ & $0.437 \pm 0.005$ & $252 $ & $-42.50 \pm 0.11$ & $33.87 \pm 0.20$ & $45.37 \pm 0.42$ \\
\hline
\end{tabular}
\newline
\end{table*}

\begin{figure}
\centering
\includegraphics[width=1.05\linewidth]{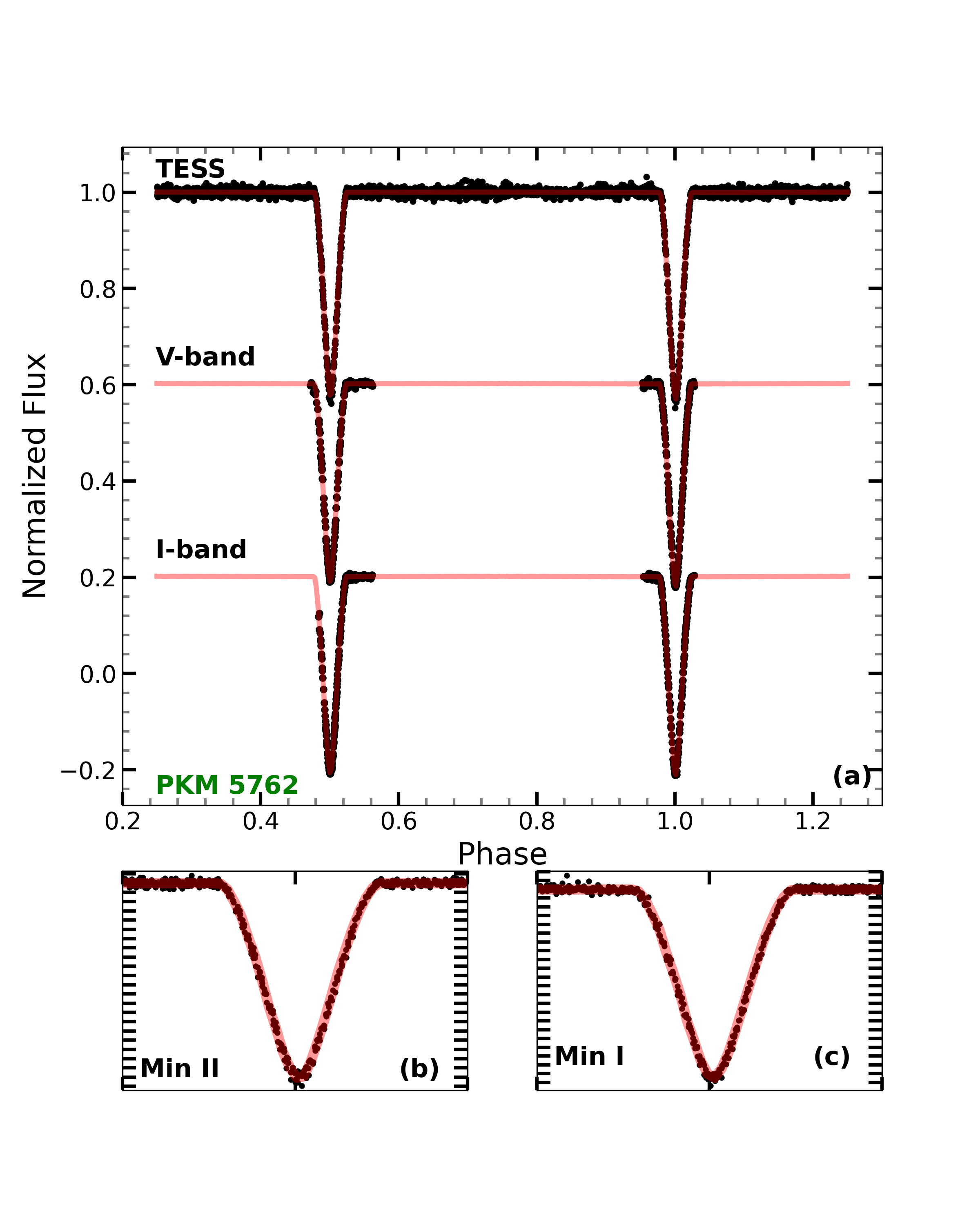}\\
\vspace*{-8mm}
\includegraphics[width=0.96\linewidth]{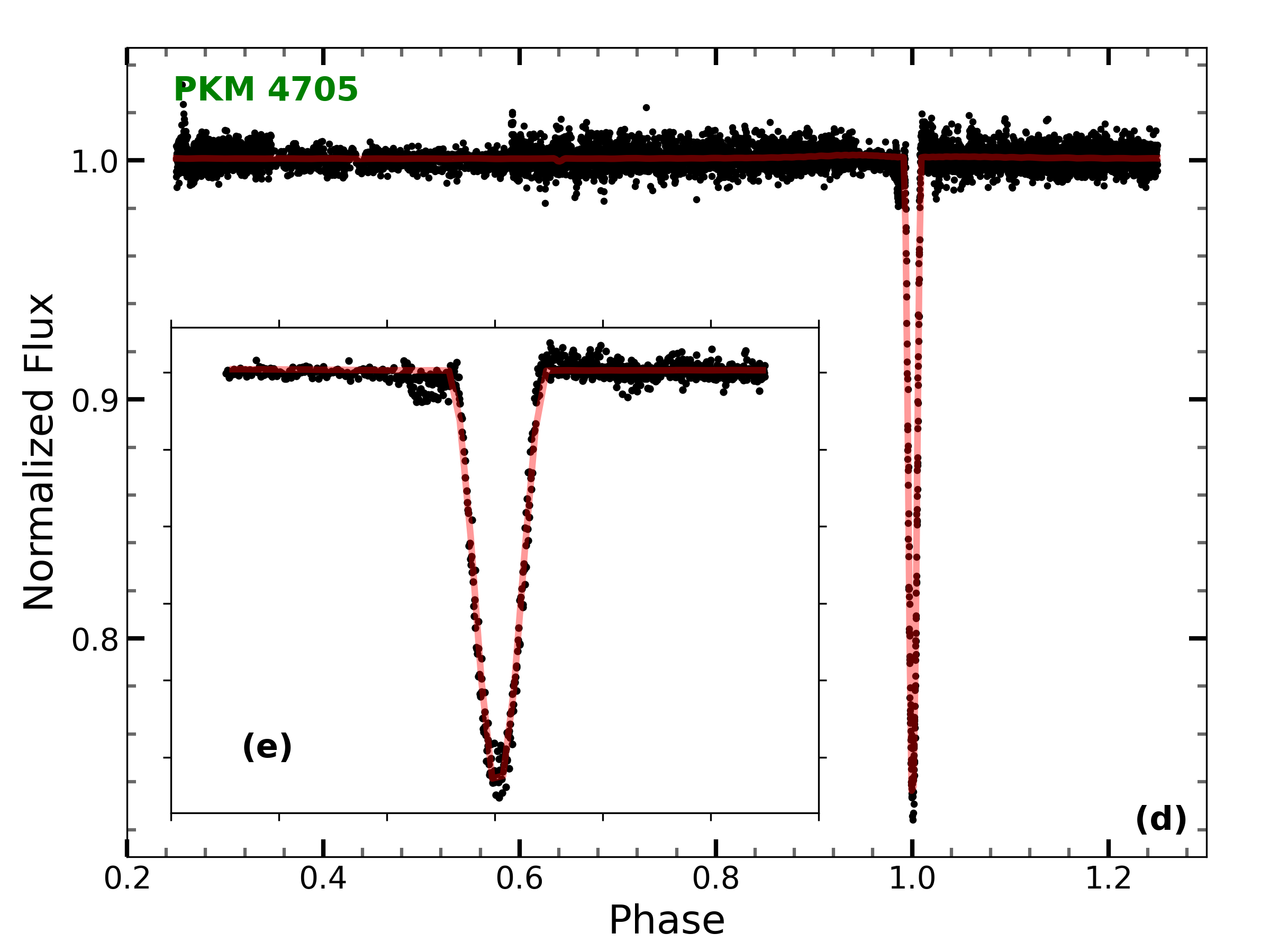}
\caption{The observed and the modelled (solid red lines) light curves of PKM 5762 (a, b, and c) and PKM 4705 (d and e).} \label{fig:LCs}
\end{figure}

During the RV and lightcurve analysis, the logarithmic limb-darkening coefficients ($x_\mathrm{1}$, $x_\mathrm{2}$) were obtained from  \citep{Claret2018A&A...618A..20C}, gravity darkening coefficients were chosen as $g_\mathrm{1}$ = $g_\mathrm{2}$ = 0.32 \citep{Lucy1967ZA.....65...89L}, and albedos were assumed to be $A_\mathrm{1}$ = $A_\mathrm{2}$ = 0.5 \citep{Rucinski1969AcA....19..245R}. Orbital period $P$, semi-major axis of the relative orbit $a$, binary center-of-mass radial velocity $V_\gamma$, orbital inclination $i$, effective temperature of the secondary component $T_\mathrm{2}$, luminosity of the primary component $L_\mathrm{1}$, the potentials of the Roche equipotential surfaces ($\Omega_1$, $\Omega_2$), mass ratio $q$, eccentricity $e$, and the longitude of periastron of the eclipsing binary, $\omega$, were adjustable parameters. After many iterations, solutions were obtained for both systems individually and the results are given in Table~\ref{tab:LCRV_results} for PKM 5762 and PKM 4705, including the uncertainties. The comparison of the synthetic models (continuous lines) obtained using these results with the observations (dots) is shown in Figure~\ref{fig:LCs} for both TESS photometric data sets and radial velocities. The figures are also shown zoomed in to make the minima more visible. As can be seen from these figures, there is very good agreement between the synthetic models and the observations.

Light curve analysis is used to determine the geometrical properties of the system. The high precision of the light curve allows us to determine other orbital elements more accurately, in particular the relative radius (a/R).  The combination of lightcurve and RV provides the outputs necessary to determine fundamental physical parameters such as mass (M), radius (R), luminosity (L), and ultimately the distance of the system. 
Errors arise from uncertainties in both photometric and spectroscopic measurements, which were estimated via Markov Chain Monte Carlo (MCMC) simulations using the \texttt{emcee} \citep{emcee2013} Python package during our data fitting procedure.
The MCMC method derives the uncertainties from the distributions of the posteriors for the various parameters. The standard deviation and confidence intervals are based on the properties of these distributions.

\begin{table}
\scriptsize
	\begin{center}
		\caption{Obtained fundamental parameters for PKM 5762 (V785 Cep)  and PKM 4705 (NGC~188~1116) based on simultaneous LC and RV solutions. The standard errors 1$\sigma$ in the last digit are given in parentheses.} \label{tab:LCRV_results}
		\begin{tabular}{lllllll}
			\hline
			Parameter                                           &  PKM 5762     & PKM 4705         \\
			\hline
			Initial epoch, T$_{\rm 0}$   (day)                  & 58893.3177(11) & 59406.502(13)  \\
			Period, P  (day)                                    & 6.50430116(12) & 35.178066(8)   \\
            Geometric parameters:                               &                &                  \\
			Inclination, i ${({^\circ})}$                       & 88.8(1)        & 83.2(2)         \\
			Eccentricity, e                                     & -              & 0.485(3)        \\
			Argument of periapse, $\omega$ ${({^\circ})}$       & -              & 246(4)          \\
			  $\Omega _{1}$                                       & 14.136(8)     & 18.393(17)         \\
            $\Omega _{2}$                                       & 14.165(7)     & 26.417(32)         \\
            Fractional radii of pri.                            &                &              \\ 
            $R_1/a$                                             &0.0763(3)        & 0.0711(3)     \\
            Fractional radii of sec.                            &                &                \\ 
            $R_2/a$                                             & 0.0744(2)       & 0.0291(2)       \\
            Radiative parameters:                               &                &                  \\ 
            T$_{\rm eff, \,1}$  (K)                             & 5857(24)       & 4829           \\
            T$_{\rm eff, \,2}$  (K)                             & 6019(25)       & 5676            \\
            Albedo$^*$ ($A_1,A_2$)                              & 0.6, 0.6       & 0.6, 0.6          \\
            Gravity brightening$^*$ ($g_1, g_2$)                & 0.32, 0.32     & 0.32, 0.32         \\
            Light ratios &&\\
            $(\frac{l_3}{l_1+l_2+l_3})_{\rm TESS}$(\%)          & 0.2(1)          &                  \\
            $(\frac{l_3}{l_1+l_2+l_3})_{\rm I}$(\%)             & 1.6(2)          &                  \\
            $(\frac{l_3}{l_1+l_2+l_3})_{\rm V}$(\%)             & 1.2(2)          &                 \\
            \hline
			Mass, $M_{1}$ ($\rm{M_{\odot}}$)                      & 1.099(7)        & 1.142(29)                   \\
			Radius, $R_{1}$  ($\rm{R{\odot}}$)              & 1.444(16)       & 4.194(53)                  \\
			Luminosity, $L_1$ ($\rm{L_{\odot}} $ )              & 2.36(13)        & 8.59(40)                    \\
			Surface gravity, $\log g_1$ (cgs)                   & 4.160(12)       & 3.251(21)                      \\
			Mass, $M_{2}$ ($\rm{M_{\odot}}$)                      & 1.075(7)        & 1.092(27)                      \\
			Radius, $R_{2}$  ($\rm{R_{\odot}}$)             & 1.412(11)       & 1.720(17)                      \\
			Luminosity, $L_{2}$ ($\rm{L_{\odot}} $ )              & 2.35(10)        & 2.76(19)                       \\
			Surface gravity, $\log g_{2}$ (cgs)             & 4.170(9)        & 4.006(17)                        \\
            \hline
		\end{tabular}
	\end{center}

\end{table}

\section{Joint Sed Fitting for the Six Binary SEDs}
\label{sec:SEDs}

\begin{table*}
\centering
\caption{Stellar parameters for 12 stars obtained from a joint SED analysis of six binary systems in NGC 188. The IDs in upper case indicate the label of the binary system, while lower case letters refer to the primary (a) and secondary (b) components.  The parameters related to the cluster and obtained jointly in the analysis are given at the bottom of the table.}
\label{tab:ngc188_sedresults}
\begin{tabular}{lllllll}
\hline
PKM   & ID& Mass                & Radius           & Temperature   & Luminosity      & Inclination \\
      &   & (${M_{\odot}}$)     & (${R_{\odot}}$)  & ($K$)         & (${L_{\odot}}$) & (${{^\circ}}$) \\
\hline
5762  & Aa & $1.098 \pm 0.005$ & $1.495 \pm 0.056$ & $5996 \pm  30$ &  $2.61  \pm 0.15$ & $88.7 \pm  0.1$ \\
      & Ab & $1.074 \pm 0.005$ & $1.351 \pm 0.036$ & $5991 \pm  15$ &  $2.14  \pm 0.10$ & $88.7 \pm  0.1$ \\
4705  & Ba & $1.174 \pm 0.012$ & $4.206 \pm 0.230$ & $4823 \pm  17$ &  $8.64  \pm 0.83$ & $83.5 \pm  1.4$ \\
      & Bb & $1.119 \pm 0.016$ & $1.660 \pm 0.117$ & $5948 \pm  74$ &  $3.13  \pm 0.32$ & $83.5 \pm  1.4$ \\
4986  & Ca & $1.041 \pm 0.021$ & $1.222 \pm 0.068$ & $5936 \pm  43$ &  $1.70  \pm 0.23$ & $78.1 \pm  4.6$ \\
      & Cb & $0.983 \pm 0.029$ & $1.061 \pm 0.069$ & $5781 \pm  95$ &  $1.16  \pm 0.22$ & $78.1 \pm  4.6$ \\
4506  & Da & $1.077 \pm 0.012$ & $1.368 \pm 0.061$ & $5992 \pm  21$ &  $2.20  \pm 0.20$ & $29.9 \pm  0.3$ \\
      & Db & $1.064 \pm 0.013$ & $1.308 \pm 0.055$ & $5978 \pm  23$ &  $1.99  \pm 0.18$ & $29.9 \pm  0.3$ \\
5147  & Ea & $1.014 \pm 0.011$ & $1.140 \pm 0.031$ & $5881 \pm  28$ &  $1.41  \pm 0.10$ & $45.2 \pm  0.4$ \\
      & Eb & $1.002 \pm 0.011$ & $1.108 \pm 0.030$ & $5849 \pm  32$ &  $1.30  \pm 0.10$ & $45.2 \pm  0.4$ \\
4411  & Fa & $1.055 \pm 0.009$ & $1.267 \pm 0.034$ & $5964 \pm  18$ &  $1.86  \pm 0.11$ & $86.6 \pm  3.3$ \\
      & Fb & $0.794 \pm 0.008$ & $0.759 \pm 0.009$ & $4977 \pm  45$ &  $0.32  \pm 0.02$ & $86.6 \pm  3.3$ \\
\hline 
\multicolumn{3}{c}{{Age}} & \multicolumn{3}{c}{{Distance}} & {$A_V$} \\
\multicolumn{3}{c}{(Myr)} & \multicolumn{3}{c}{(pc)} & {(mag)} \\
\hline
\multicolumn{3}{c} {$6410 \pm 270$} & \multicolumn{3}{c}{$1852 \pm 17$} & {$0.34 \pm 0.04$} \\
\hline
\hline
\end{tabular}

\textit{Note: These results were derived for an assumed metallicty of [Fe/H] = +0.025, which according to \citet{Choi2016} and \citet{Asplund2009} corresponds to $Z = 0.0150$ with $X = 0.715$ and $Y = 0.270$. The uncertainty in distance derived from this fit is small only because we set the priors on distance to match the astrometric results (see Sect. 5). However, in other fits we explored a range of [Fe/H] values, and also opened up the prior on the distance. Our independently determined photometric distance is $1897 \pm 58$ pc.}
\end{table*}

We start by collecting the available archival spectral flux measurements from 0.19 to 11.6 microns for each of the six selected binaries.  We access these data from VizieR SED\footnote{A.-C. Simon \& T. Boch: \url{http://vizier.cds.unistra.fr/vizier/sed/}} \citep{ochsenbein00} which, in turn, utilizes systematic sky coverage of such surveys such as Pan-STARRS \citep{chambers16}, SDSS \citep{gunn98}, 2MASS \citep{2MASS}, WISE \citep{WISE}, Galex \citep{bianchi17}, and Swift (UVOT; \cite{Poole2008}). For each of the six binaries there are typically 30-33 spectral points that we select. We fit the six SEDs of our selected binaries jointly under the assumption that all 12 stars were born at the same time (or at least within a common 20 Myr timespan), and that there was no mass transfer within any of the binaries.  

There were 21 free (i.e., fitted) parameters: the 12 stellar masses, 6 orbital inclination angles, and a common age of NGC 188), distance, and interstellar extinction, $A_V$.  The formal input data that enable these parameters to be determined are the $\sim$180 SED points, 12 $K$ velocities (taken from Table \ref{tab:binaries}, and either the use of the Gaia distance (see Sec. 5) or not.  Additionally, we utilize the MIST evolution tracks (\citealt{choi16}; \citealt{dotter16}; \citealt{paxton11}; \citealt{paxton15}; \citealt{paxton19}) to connect each mass, and a common age, with its radius and $T_{\rm eff}$.  The orbital periods and eccentricities are known well enough (see Table \ref{tab:binaries}) that they are simply fixed at their measured values during the fit. In all we ran the joint SED fits for six different cases of metallicity in the MIST evolution tracks: [Fe/H]= $-0.1$, $-0.05$, 0.0, $+0.025$, $+0.05$, and $+0.1$. The stellar atmosphere's models were taken from \citet{castelli03} and fixed at solar composition since the established metallicity of this cluster is  measured to be sufficiently close to solar (see, Table \ref{tab:ngc188_metallicity}, and references therein) for this latter purpose.

The fits were carried out with an MCMC code (\citealt{ford05}; \citealt{rappaport21}).  We ran ten MCMC chains each with between 200 and 500 million links.  The run time on a current laptop was 1.5-4 hours per chain.  Here we mention a few minor technical notes associated with the fit.  First, we take the two stellar masses in a given binary, coupled with the orbital period and the eccentricity, to compute the two expected $K$ velocities.  These are then compared to the measured $K$ velocities and the resultant contribution to $\chi^2$ was added to the $\chi^2$ value from the SED fit.  Second, regarding the SED points, we somewhat arbitrarily assigned an 8\% uncertainty to each point regardless of the stated uncertainties that are generally much smaller.  Our estimate is based on the typical empirical scatter from point to point.  In some cases this scatter arises from eclipses, but only two of our six binaries are eclipsing.  However, we suspect that there are other uncertainties than the formal statistical ones that are cited.  Third, for the SED points provided by Swift's UVOT at 0.193 microns, we use a 50\% error bar because these points do not fit well.  For the WISE 11.6 micron point we typically use a 25\%  error because this point is quite uncertain for the NGC 188 stars.    

\begin{figure*}
\centering
\includegraphics[width=0.32\textwidth]{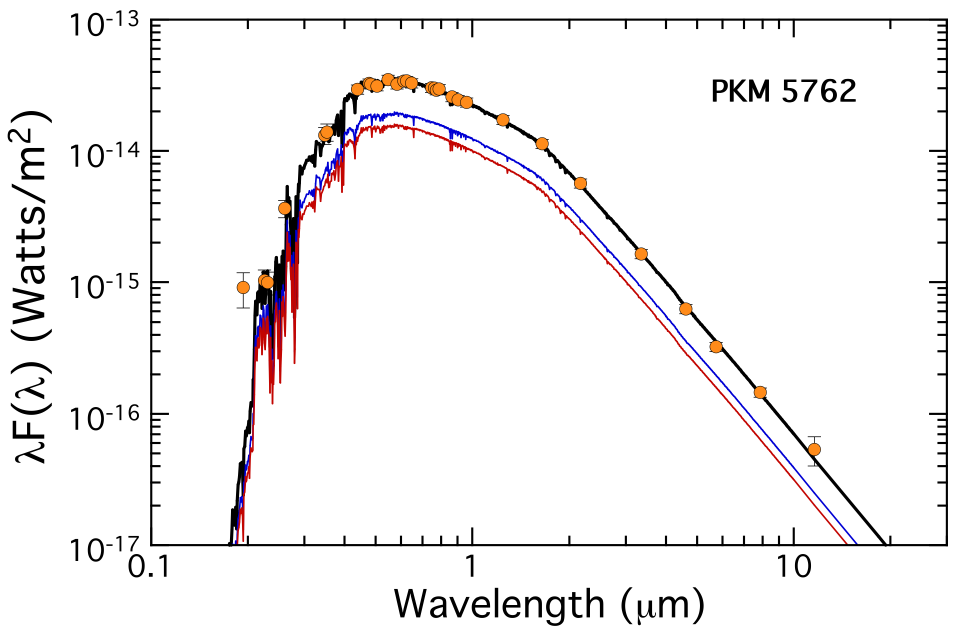}
\includegraphics[width=0.32\textwidth]{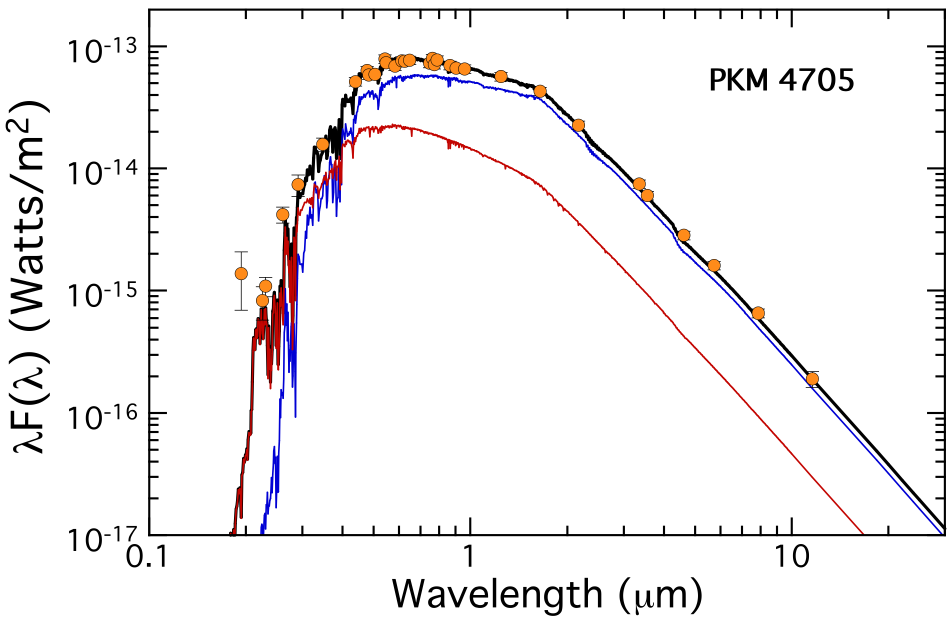}
\includegraphics[width=0.32\textwidth]{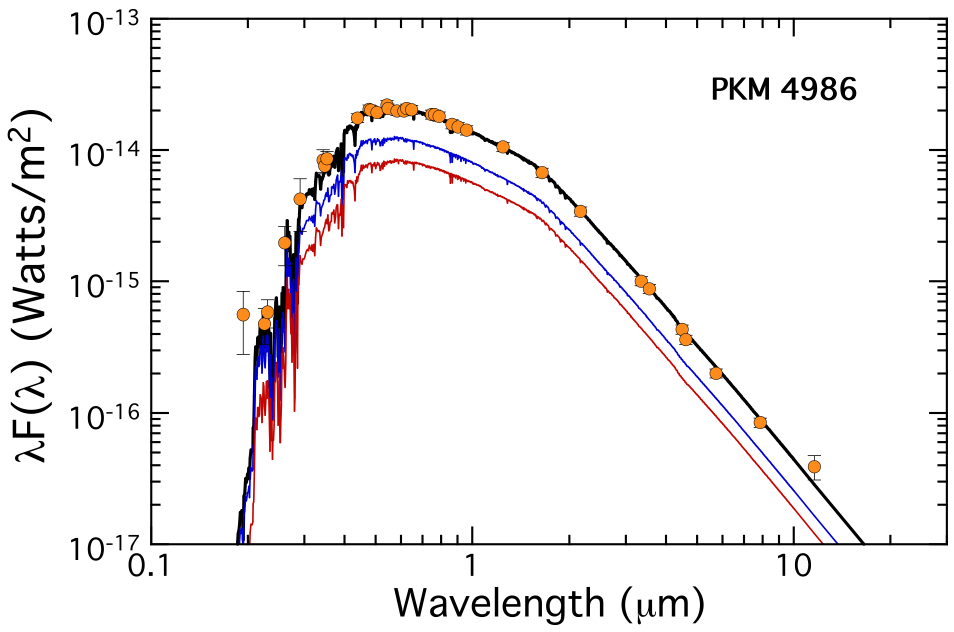}
\includegraphics[width=0.32\textwidth]{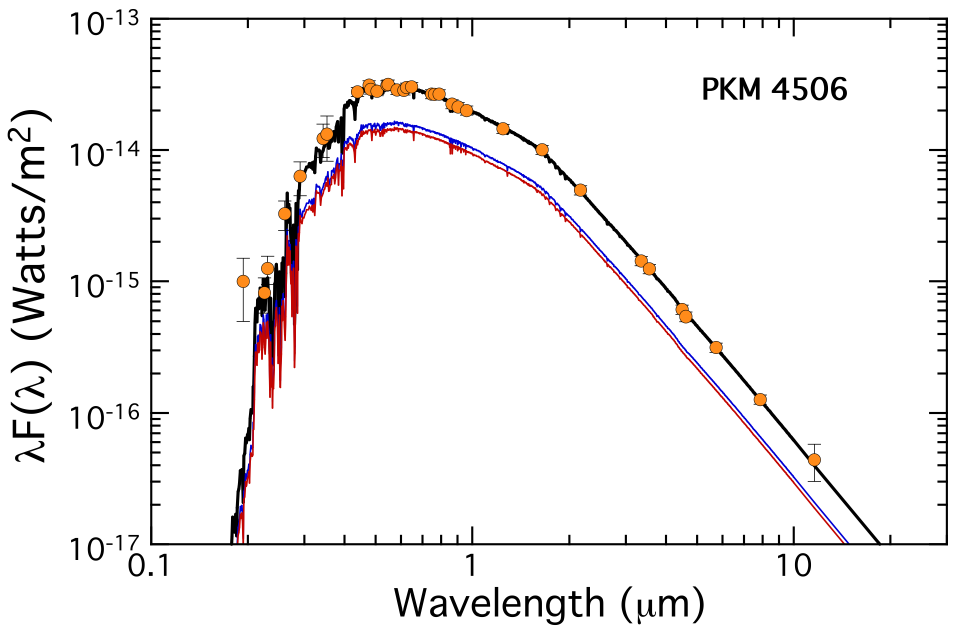}
\includegraphics[width=0.32\textwidth]{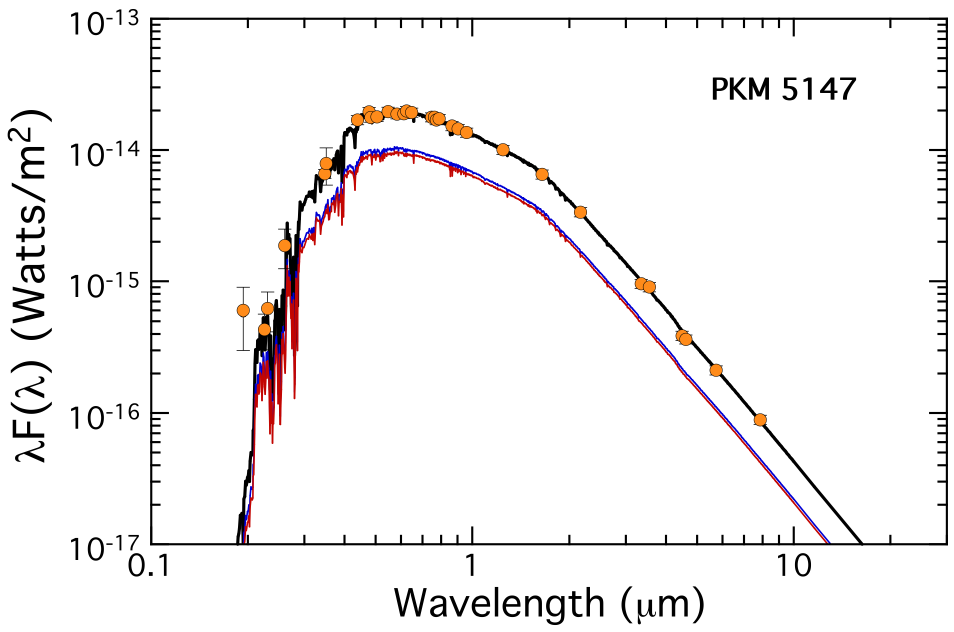}
\includegraphics[width=0.32\textwidth]{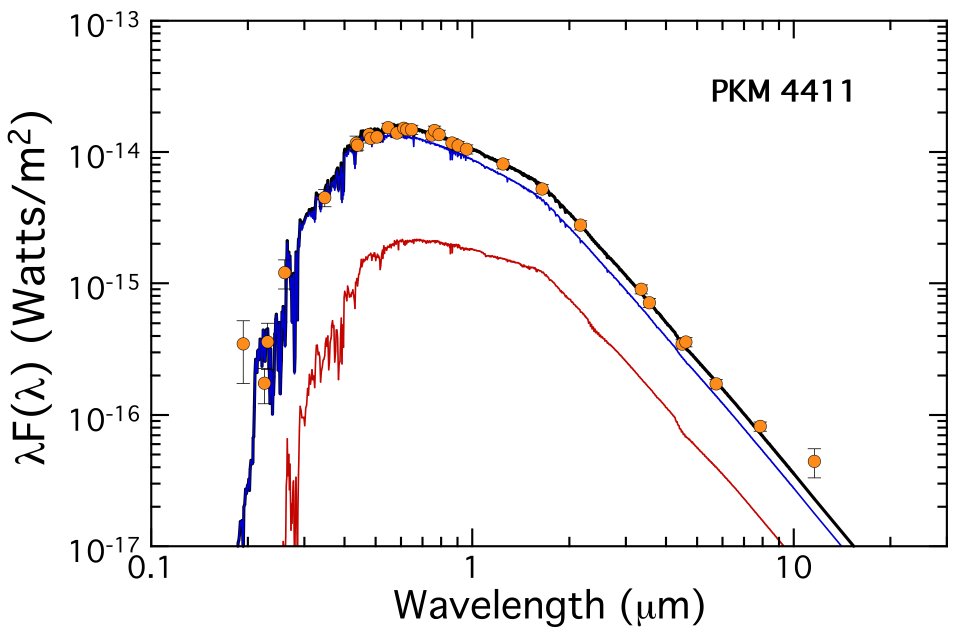}
\caption{The SEDs of six binaries that were fit jointly with a common age, distance, and $A_V$. A composition of [Fe/H] = +0.025 was adopted for this fit.  Figure \ref{fig:age_Fe} shows how the age varies with [Fe/H]. The blue and red curves are model fits for the individual stars while the black curve is the composite SED.  The data points are represented by the orange circles with error bars. }
\label{fig:seds}
\end{figure*}

The results are shown in Figure \ref{fig:seds}, for [Fe/H] = +0.025, as a six-panel montage of SED fits.  The vertical scales on all six panels are set to the same range except for the upper right panel that extends to a factor of 2 higher in flux.  The fitted parameter results, i.e., the 12 stellar masses, radii, $T_{\rm eff}$'s, and luminosities, as well as the orbital inclinations, distance to the cluster and its age, and the extinction are all listed in Table \ref{tab:ngc188_sedresults}.

\begin{figure}
\centering
\includegraphics[width=0.99\columnwidth]{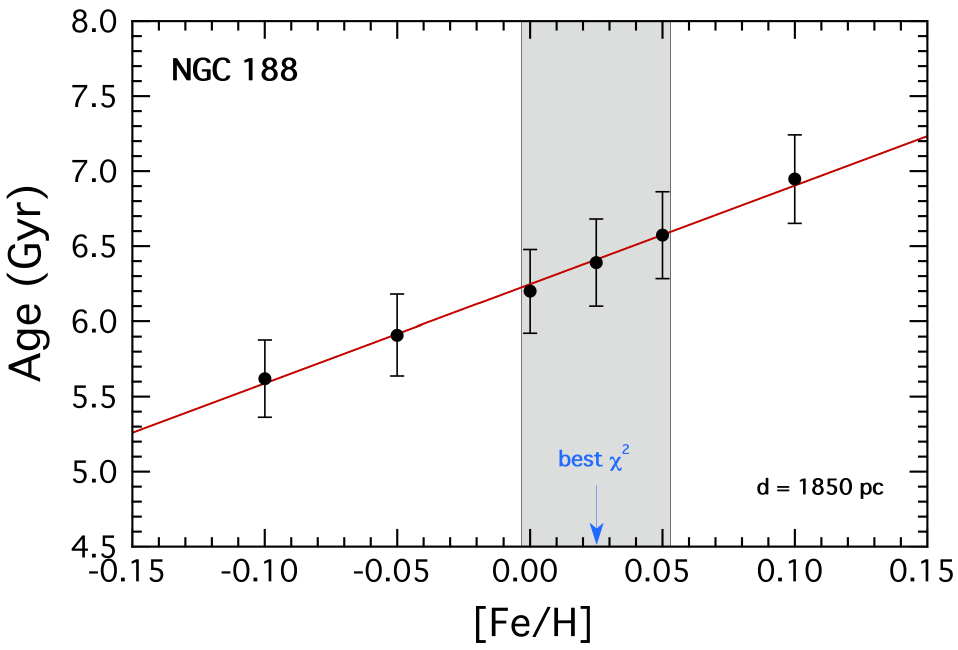}
\caption{Age of NGC 188 vs.~the assumed metallicity [Fe/H]. The same type of SED fits used to produce Table \ref{tab:ngc188_sedresults} and Fig.~\ref{fig:seds} was carried out for five additional metallicities of [Fe/H] = $-0.1$, $-0.05$, 0.00, +0.05, and +0.1. We find a linear relation between the age of NGC 188 and metallicity given by Eqn.~\ref{eqn:age_fe}. The point at [Fe/H]=+0.025 represents the average metallicity found in the literature, while the arrow at [Fe/H]=+0.025 marks the metallicity yielding the best $\chi^2$ fit to the SEDs. The shaded area represents the approximate $\pm 1\sigma$ uncertainty region for [Fe/H] as measured spectroscopically (see Table.~\ref{tab:ngc188_metallicity})}
\label{fig:age_Fe}
\end{figure}

In Figure \ref{fig:age_Fe} we show just the age emerging from the SED fits as the metallicity is varied from [Fe/H] of $-0.1$ to +0.1. The results can be summarized with the following linear relation:
\begin{equation}
{\rm age} \simeq (6.25 \pm 0.28)+(6.6 \pm 1.7) [\rm{Fe/H}]~{\rm Gyr}
\label{eqn:age_fe}
\end{equation}
This expression implies that the most likely age, for a most probable [Fe/H] =+0.025, is 
$${\rm age} ~ \simeq 6.41 \pm 0.28 \pm 0.18 ~{\rm Gyr}$$ where the first error bar is a statistical uncertainty and the second is the uncertainty in age due to the uncertainty in the cluster metallicity. Thus, the age is distinctly correlated with metallicity, [Fe/H].  However, we do not make metallicity a fitted parameter because the {\it quality} of the fit (i.e., as measured by $\chi^2$), is not strongly dependent on the age.  Therefore, the metallicity is better determined by precision spectroscopy observations.

If we open up the prior on the distance to NGC 188 (as opposed to taking the narrow prior of $1850 \pm 40$ pc used to generate Table \ref{tab:ngc188_sedresults}) we find an independent, but less precise, photometric distance to NGC 188.  This distance is modestly correlated with age.  The correlation is:
$$d_{\rm ph} = 1897 + 175 \,({\rm age}-6.41 \,{\rm Gyr}) ~~{\rm pc} $$ with a correlation coefficent of 0.73.
For an uncertainty in the age of 0.3 Gyr, the independent photometric distance is 
 $$d_{\rm ph} = 1897 \pm 58 \,{\rm Gyr} ~~{\rm pc} $$
This photometric distance is quite consistent with the one found from the analysis of Gaia data for this cluster (see Sect. \ref{sec:astrometric}), and which we use in the final SED analysis $d_{\rm astr} = 1850 \pm 40$\,pc). 

We also find the cluster age is not correlated with $A_V$ even when we open up the priors on the distance.

Finally, we note that the precision on the age results is in no small part due to the binary PKM 4705. Without this binary included, the age determination would be $6.0 \pm 0.5$ Gyr. However, the full set of results we obtain also rests in a significant way on the other five binaries which contain eight stars that are distinctly evolved away from the ZAMS. The addition of these other five binaries to the collection adds a certain robustness to the age determination.

\section{Astrometric Analysis}
\label{sec:astrometric}

Recent advancements in astrometric surveys, particularly those provided by the Gaia mission, have revolutionized the identification of stellar cluster members. Contemporary studies increasingly utilize advanced techniques—such as probabilistic approaches, proper motion vector point diagram (VPD) analyses, machine learning algorithms, and clustering in multi-dimensional parameter spaces—to compute stellar membership probabilities with unprecedented precision \citep[e.g.,][]{CantatGaudin2018}.
Constructing accurate and comprehensive membership lists is essential not only for deriving the fundamental astrophysical parameters of star clusters (e.g., age, metallicity, mass function), but also for probing their internal dynamical structures (e.g., mass segregation, binary fraction, velocity dispersion) and tracing their orbital evolution within the Galaxy.
In this context, the membership catalog and kinematic structure of NGC 188—recognized as one of the oldest and most thoroughly investigated open clusters in the Milky Way—have undergone significant refinement in recent years, largely due to the availability of high-precision Gaia data \citep[e.g.,][]{Platais2003, Geller2021}. As a long-standing benchmark in stellar evolution studies, NGC 188 continues to serve as a critical reference for testing theoretical models and understanding the long-term dynamical evolution of open clusters.

\begin{figure}
\centering
\includegraphics[width=1.00\linewidth]{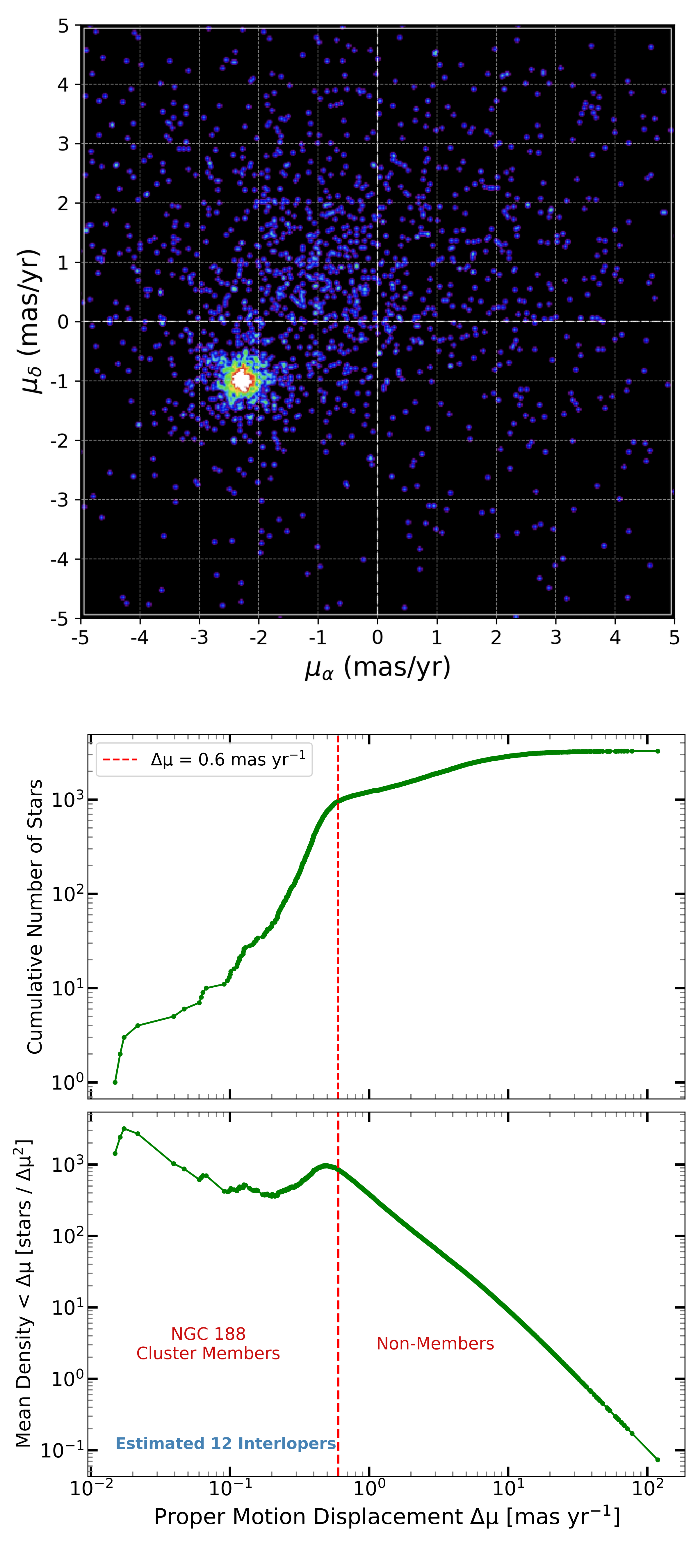}
\caption{\textbf{Top panel:} Proper motion density distribution of stars in the NGC 188 field, centered at (0,0) based on Gaia DR3 data. The dotted horizontal and vertical lines indicate $\mu_\alpha = 0$ and $\mu_\delta = 0$, defining the coordinate reference frame used in this visualization. \textbf{Middle panel:} Cumulative distribution of proper motion displacements $\Delta\mu$ from the cluster center. The red dashed vertical line marks the threshold value of $\Delta\mu = 0.6$ mas~yr$^{-1}$, used to distinguish cluster members from field stars.   \textbf{Bottom panel:} Mean cumulative density (number of stars per $\Delta \mu^2$) as a function of $\Delta\mu$. This plot helps visualize how sharply the density of stars drops beyond the selected membership threshold, reinforcing the distinction between cluster members and non-members.}
\label{fig:pm_density_ngc188}
\end{figure}

To determine stellar cluster membership in the region of the open cluster NGC 188, we retrieved astrometric data from the Gaia DR3 catalogue using the Gaia Archive interface \citep{GaiaCollaboration2021}. We extracted 3,017 stars from the Gaia DR3 catalogue within a cone of radius $15’$ centered on NGC 188. The radius of the cluster on the sky is given by \cite{Sanchez2018} as $15'$, but it is difficult to discern a sharp spatial contrast that clearly delineates the extent of NGC 188 on the sky. We find, however, that cluster membership is much better defined in proper motion (PM) space.  The top panel in Fig.~\ref{fig:pm_density_ngc188} shows all the Gaia stars in a PM region $|\Delta \mu_\alpha| < 5$ and $|\Delta \mu_\delta| < 5$ mas yr$^{-1}$.  The image is in pseudo-color, with color qualitatively indicating the stellar density in proper motion space.  The dense concentration centered at $\mu_\alpha = -2.5$~mas\,yr$^{-1}$ and $\mu_\delta = -1.1$~mas\,yr$^{-1}$, clearly indicates the cluster NGC 188.

To make the cluster membership more quantitative, we show in the middle panel of Fig.~\ref{fig:pm_density_ngc188} the cumulative number of stars as a function of the displacement in proper motion space from the center of the concentration seen in the top panel (at $\mu_\alpha = -2.5$~mas\,yr$^{-1}$, $\mu_\delta = -1.1$~mas\,yr$^{-1}$).  The number rises steeply even in a log-log plot, but then abruptly changes to a much lesser slope at $\Delta \mu \simeq 0.6$ mas yr$^{-1}$.  This implies that the density of stars also changes abruptly near this displacement.  This is quantified in the bottom panel of Fig.~\ref{fig:pm_density_ngc188} where we show the mean star density (stars per $\Delta \mu^2$) within $< \Delta \mu$ as a function of displacement in mas yr$^{-1}$.  This shows how the mean density is approximately constant (to within a factor of $\sim 2$), and then the local density must drop suddenly beyond that displacement in order for the mean density to  fall off approximately as $\Delta \mu^{-3/2}$ after a displacement of $\simeq 0.6$ mas yr$^{-1}$. (If the mean density in PM space fell like $\Delta \mu^{-2}$, this would imply zero external star density.) We take $\Delta \mu = 0.6$ mas yr$^{-1}$ as our working definition of the cluster boundary in PM space. 

We also eliminate all stars that have a Gaia RUWE $> 1.4$ which indicates a binary (or higher-order system) that is sufficiently wide to adversely affect the Gaia distance measurement.  With that one additional restriction, we find that there are 333 stars within the above specified PM region with which we have to work. These have a very high confidence ($> 90\%$, see calculation below) of being actual cluster members.  After applying a nominal zero point correction to the parallaxes of these stars of 0.017 mas \citep{Lindegren2021}, we find that the median parallax is 0.5405 mas, and the median distance is 1850 pc, with an rms scatter of 170 pc. 

To estimate the number of false positive cluster members within the specified PM displacement, i.e., the number of `interlopers', we compute the density of field stars in this neighborhood to be the total number of stars within an area of 150 square mas yr$^{-1}$.  Beyond such an area, the density of all stars in PM space at this RA and Dec on the sky falls to zero.  Thus, the mean density of stars in PM space is  $\simeq 3000/150 \simeq 20 \, ({\rm mas~yr}^{-1})^{-2}$.  Since we have chosen a maximum $\Delta \mu$ displacement of 0.6 mas yr$^{-1}$ this implies a total estimated number of  interloper stars among our selected cluster stars of $\simeq 23$. 

Then the question becomes -- with a collection of 333 total stars with a median measured parallax of 1850 pc, an rms scatter of 170 pc, and standard error of 9 pc, what is the additional uncertainty introduced by some 23 interloper stars?  We have simulated $10^6$ random realizations of this scenario using Monte Carlo techniques, and empirically find that we should more realistically take the standard error to be  12 pc due to the interloper stars.  Moreover, we find from these MC simulations that the median value of the distance, including the interlopers, is unchanged by more than a couple of pc even when we increase the mean and rms distance of the interlopers to be 1950 pc and 425 pc, respectively.

\begin{figure}
\centering
\includegraphics[width=1.00\linewidth ]{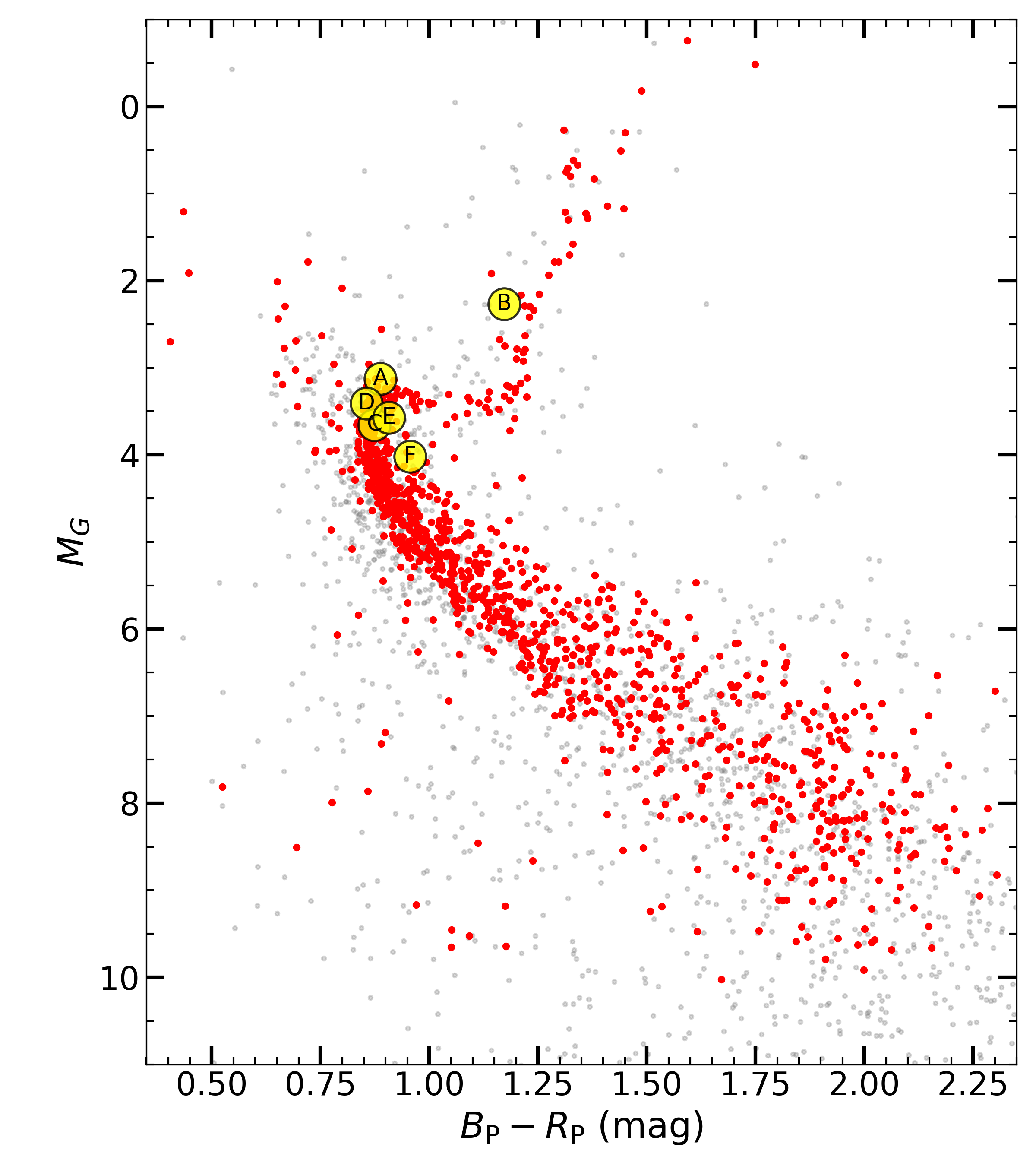} 
\caption{Hertzsprung–Russell diagram of the cluster is plotted using Gaia photometry, where cluster members are shown in red and the target binary systems are highlighted and labeled at their respective positions.} 
\label{fig:gaiaDR3HR}
\end{figure}

Figure~\ref{fig:gaiaDR3HR} shows the Gaia-based Hertzsprung–Russell (H-R) diagram for stars with high-confidence cluster membership based on proper motion selection. The color–magnitude distribution reveals a clearly defined main sequence, a sparsely populated subgiant branch, and a red giant clump consistent with the expected evolutionary features of an old open cluster like NGC 188. The six spectroscopic binaries analyzed in this study are highlighted individually in the diagram. Their positions generally align with the cluster sequence, after considering that most of the stars in the cluster are single, and our objects are binaries. This overall consistency supports the validity of the adopted membership threshold in proper motion space and affirms that the selected stars represent a physically coherent population.

\section{Results and Conclusion}
\label{sec:results}

Through the combined analysis of radial velocity (RV) curves and spectral energy distributions (SEDs) for six double-lined binary systems in NGC 188, we have obtained accurate stellar and orbital parameters for twelve individual stars. The masses and $T_{\rm eff}$ values are typically determined to accuracies of $\sim$1.3\% and $\sim$1\%, respectively, while the stellar radii are good to $\sim$4\%.
By using the fact that all twelve stars analyzed in this study belong to the same cluster and therefore share common properties such as age, distance, and chemical composition, it is possible to adopt a consistent evolutionary framework for interpreting their physical parameters. 

The astrometric results are visualized in Figure~\ref{fig:pm_density_ngc188}, which illustrates the proper motion density and the membership threshold used to isolate cluster stars from the field population. The steep density contrast between the inner and outer proper motion regions is further confirmed by the cumulative mean density profile shown in the bottom panel of Figure~\ref{fig:pm_density_ngc188}. Based on the estimated background density, approximately 23 field interlopers (only 7\% of the total) are expected within the adopted selection radius of $\Delta\mu \leq 0.6$~mas~yr$^{-1}$, confirming the robustness of our kinematic membership boundary. We have shown via Monte Carlo simulation that $\sim$23 interlopers among 333 stars makes a negligible difference in the determination of the median value of the cluster distance. 
 
Using the approach of jointly fitting the SEDs of six binaries and the measured $K$ velocities of each star, we were able to determine the age of NGC 188 as $6.41 \pm 0.28$ Gyr (statistical uncertainty).  A significant portion of that high accuracy is due to the binary PKM 4705 which contains a highly evolved star.  However, we point out that, even without this particular binary, we are still able to determine the age as 6.0 +/- 0.5 Gyr, thereby emphasizing the utility of this method for age determination.  However, it clearly is beneficial to use binaries with as evolved components as are available. 

\begin{figure}
\centering
\includegraphics[width=0.45\textwidth]{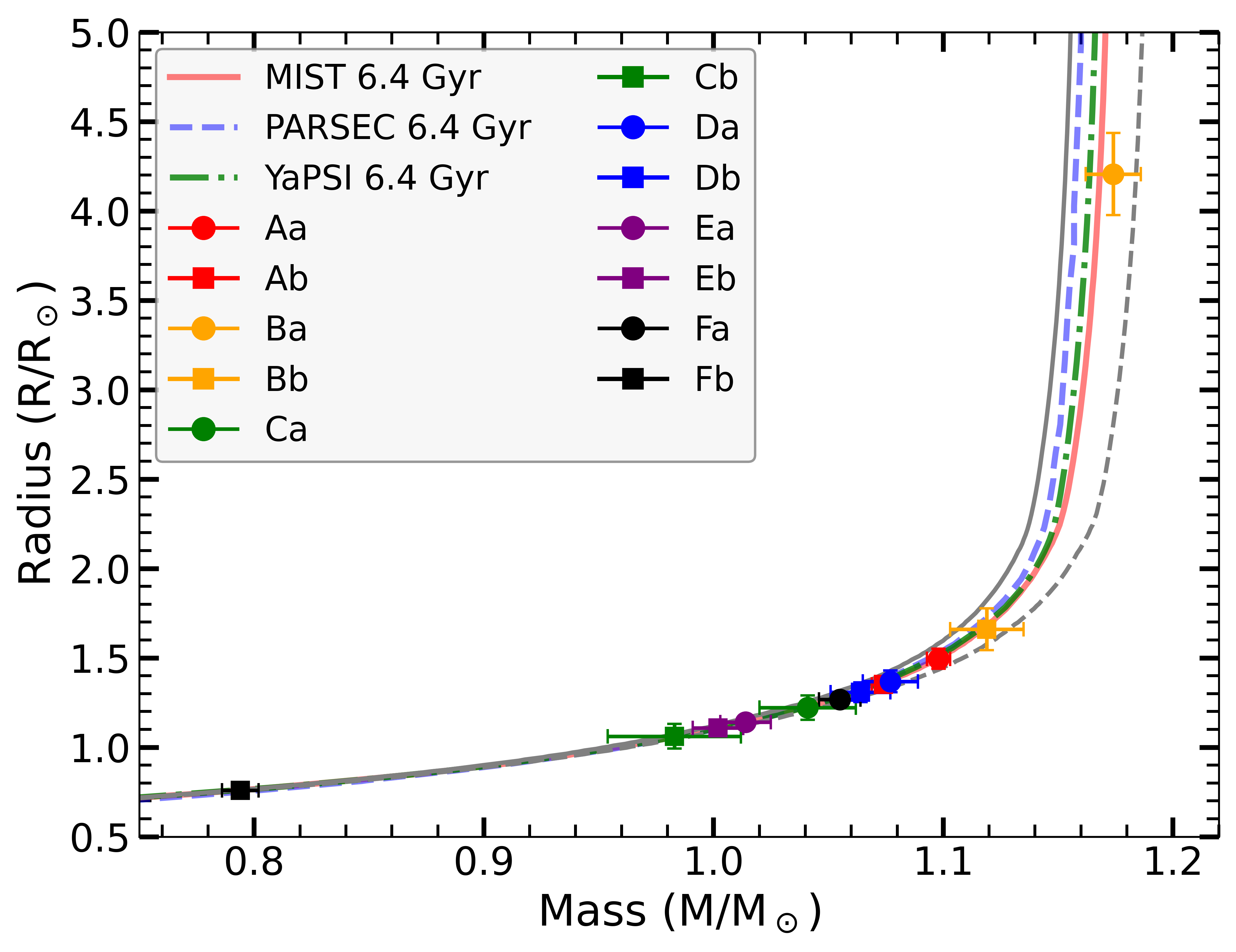}
\includegraphics[width=0.45\textwidth]{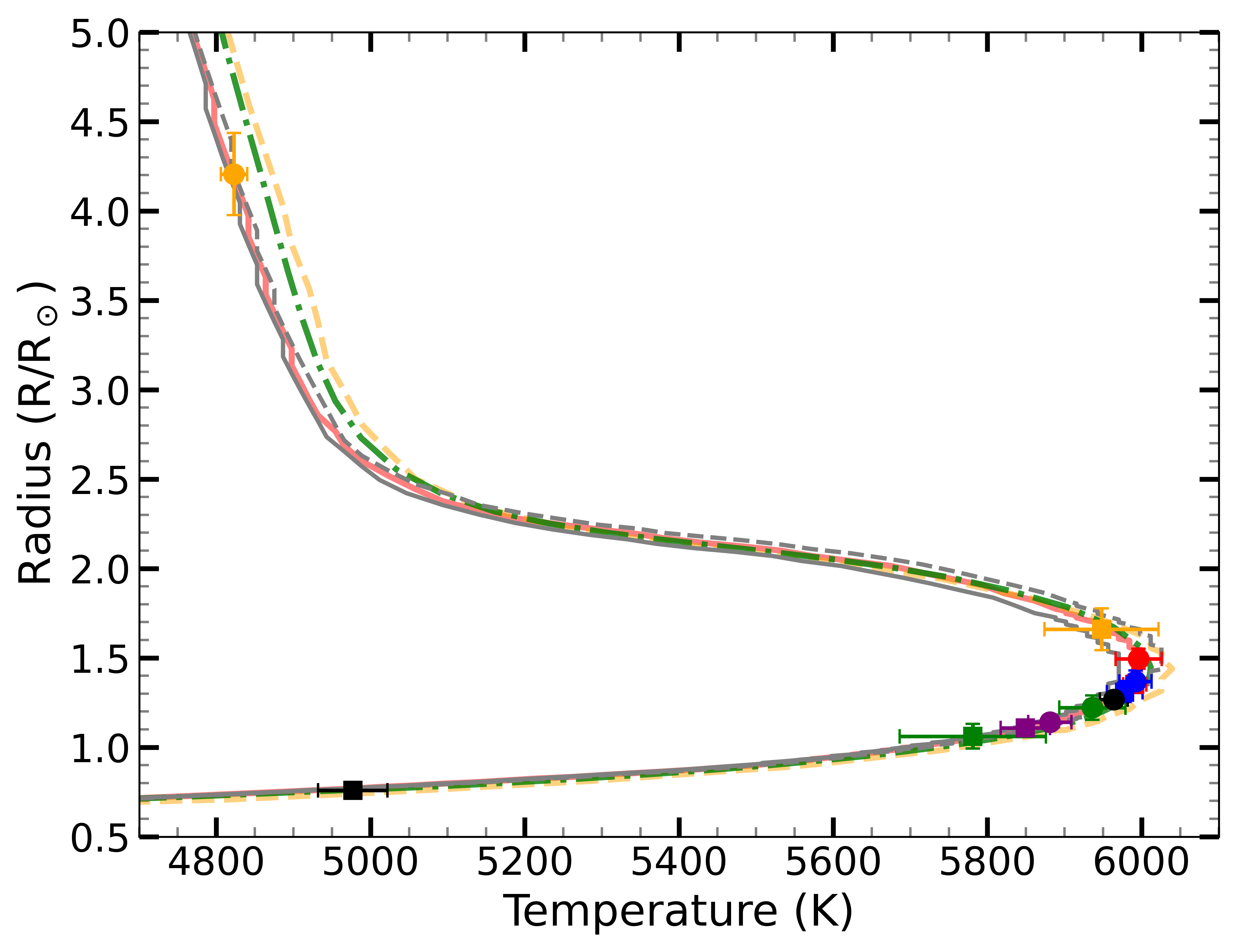}
\includegraphics[width=0.48\textwidth]{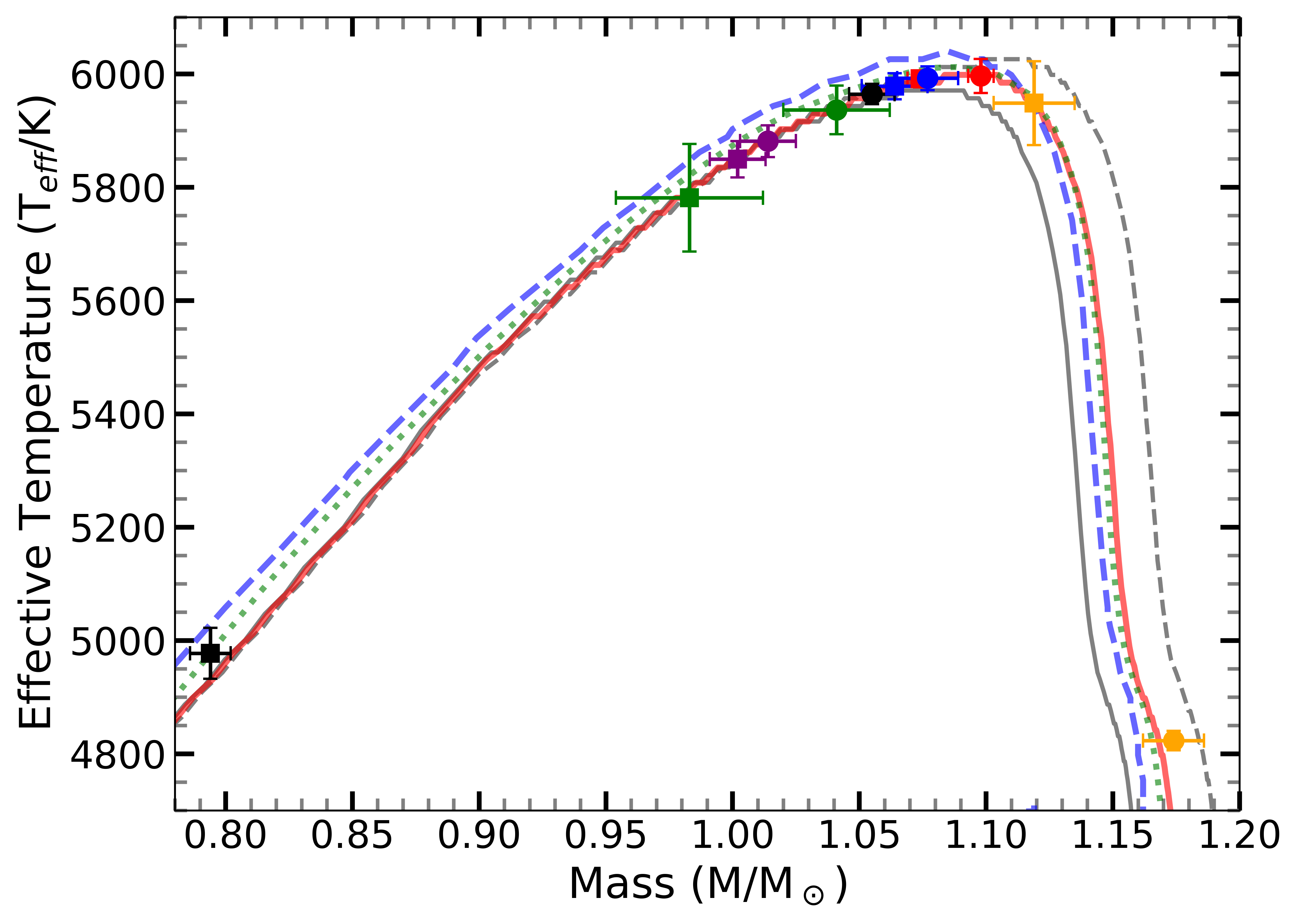} 
\caption{R-M, R-T and T-M planes of six binary systems that are members of the NGC 188 cluster. MIST (solid red), PARSEC (blue dash) and YaPSI (Y2) green dash) isochrones with ages of $6.4\pm0.3$ Gyr are also shown. MIST isochrones of 6.1 Gyr (dashed black line) and  6.7 Gyr (continuous black line) are also shown to give an idea of the range of uncertainty. The yellow circles indicate the six binary systems analyzed in this study.  The IDs of the binary systems A, B, C, D, E, and F are given in Table~\ref{tab:ngc188_sedresults}.}
\label{fig:iso}
\end{figure}

Using the derived fundamental parameters - masses, radii, effective temperatures and luminosities - for each binary component, we assess the evolutionary states of the stars by comparing their positions on various diagnostic diagrams, including the temperature-radius (T-R), mass-radius (M-R) and mass-temperature (M-T) planes. These diagrams, shown in Figure~\ref{fig:iso}, incorporate stellar evolution tracks and isochrones from three well-established models: PARSEC \citep{Bressan2012}, MIST\footnote{Since the MIST evolution tracks were used in fitting the SEDs, naturally, we expect the corresponding isochrones to match the stellar masses, radii, and $T_{\rm eff}$.  However, comparison with other independently computed model isochrones illustrates the robustness of our approach.} (\citealt{choi16}; \citealt{dotter16}; \citealt{paxton11}; \citealt{paxton15}; \citealt{paxton19}), and Y2 (YaPSI) \citep{Demarque2004,Spada2017}. The plots also show the best-fitting isochrone for the cluster, corresponding to an age of 6.4~Gyr, along with bounding isochrones at 6.1 and 6.7~Gyr, representing the $\pm1\sigma$ uncertainty range.  Previous studies of the open cluster NGC 188 have yielded a range of estimates for its fundamental parameters. As summarized in Table~\ref{tab:ngc188_metallicity}, recent age determinations for the cluster span approximately 1 Gyr, while derived metallicities ([Fe/H]) also exhibit considerable dispersion. This work contributes to refining these parameters, presenting an analysis that reduces the scatter in the published age and metallicity values for NGC 188.

As can be seen in the plots, the positions of all twelve stars from the six selected binary systems show a high degree of agreement with the 6.4~Gyr isochrone, with most components falling well within the $\pm1\sigma$ bounds. This agreement strengthens the reliability of our derived stellar parameters and supports the consistency of the age estimate, which is constrained by various observational datasets and the adopted modelling approach. The coherence observed across all diagrams reinforces the validity of using these binaries as age tracers for the cluster and highlights the precision of the combined astrometric, photometric and spectroscopic analyses employed in this study.

The astrometric analysis, based on Gaia DR3 proper motion and parallax data, enabled a robust membership determination for 1,175 candidate stars in the cluster. The refined proper motion values ($\mu_\alpha = -2.5 \pm 0.12 $ mas\,yr$^{-1}$, $\mu_\delta = -1.1 \pm 0.12$ mas\,yr$^{-1}$) and distance estimate ($1850 \pm 7$ pc) are in excellent agreement with independent literature values but offer enhanced precision. The combination of photometric, spectroscopic, and astrometric techniques in this work exemplifies a multi-pronged approach to cluster astrophysics. Future improvements in astrometric precision (e.g., Gaia DR4) and SED modeling accuracy could further refine the inferred evolutionary properties of binary systems in benchmark open clusters like NGC 188.

\section*{Acknowledgements}
We thank an anonymous referee for some very constructive suggestions for improving the manuscript. This study is based on data collected by the TESS mission, funded by the NASA Explorer Program and made publicly available by the TESS Science Office through the Mikulski Archive for Space Telescopes (MAST). This study was supported by the Scientific and Technological Research Council of Turkey (T\"UB\.ITAK 112T766 and 117F188). The numerical calculations reported in this paper were fully/partially performed at T\"UB\.ITAK ULAKBIM, High Performance and Grid Computing Center (TRUBA resources). KY thanks Churchill College for his fellowship.

\section*{Data Availability}

The TESS observations used in the analysis of binary light curves are available online to the public through the Mikulski Space Telescope Archive (MAST). If desired, the data used in this study can be obtained from the MAST servers or requested from the authors.



\bibliographystyle{mnras}
\bibliography{ngc188} 



\bsp	
\label{lastpage}
\end{document}